\documentclass[preprint]{imsart}
\RequirePackage{amsthm,amsmath,amsfonts,amssymb}
\RequirePackage[authoryear]{natbib}
\RequirePackage[colorlinks,citecolor=blue,urlcolor=blue]{hyperref}
\RequirePackage{graphicx}

\startlocaldefs
\newtheorem {Prop}{Proposition} [section]

 \newtheorem {Nota}{Remark}[Prop] 
 \newcommand{\Rea}{\mbox{$\mathbb{R}$}}

\endlocaldefs

\begin{document}

\begin{frontmatter}
\title{Statistical Depth based Normalization and Outlier Detection of Gene Expression Data}
\runtitle{Depth Based Normalization and Outlier Detection}

\begin{aug}
\author[A]{\fnms{Alicia} \snm{Nieto-Reyes}\thanksref{t1}\ead[label=e1]{alicia.nieto@unican.es}}
\and
\author[B]{\fnms{Javier} \snm{Cabrera}\thanksref{t2}\ead[label=e2]{cabrera@stat.rutgers.edu}}

\address[A]{Department of Mathematics, Statistics and Computer Science,\\
University of Cantabria, Spain,
\printead{e1}}
\address[B]{Department of Statistics, Rutgers University, USA,
\printead{e2}}
\thankstext{t1}{Partially supported by the Spanish Ministry of Science, Innovation and Universities  MTM2017-86061-C2-2P}
  \thankstext{t2}{Supported by NPR 07-1045-2-395 grant from Qatar National Research Fund (QNRF) and by a Rutgers Newark Chancellor seed grant}


\end{aug}


\begin{abstract}
Normalization and outlier detection belong to the preprocessing of gene expression data. We propose a  natural  normalization procedure based on statistical data depth which normalizes to the  distribution of gene expressions of the most representative gene expression of the group. This differ from the standard method of quantile normalization, based on the coordinate-wise median array that lacks of the well-known properties  of the one-dimensional median. The statistical data depth maintains those good properties. Gene expression data are known for containing outliers. Although detecting outlier genes in a given gene expression dataset has  been broadly studied, these methodologies do not apply for  detecting outlier samples, given the difficulties posed by the high dimensionality but low sample size structure of the data. The standard procedures used for  detecting outlier samples are visual and based on dimension reduction techniques; instances are multidimensional scaling and spectral map plots. For detecting outlier genes in a given gene expression dataset, we propose an analytical procedure and based on the Tukey's concept of outlier and the notion of statistical depth, as previous methodologies lead to unassertive and wrongful outliers. 
We reveal the outliers of four datasets; as a necessary step for further research.
\end{abstract}

\end{frontmatter}

\section{Introduction}\label{intro}

The study of gene expressions was first pursued through the use of DNA microarray and, although it is still in use, the technology is being replaced by the RNA sequencing (RNAseq).
Both techniques offer the possibility to monitor the behavior patterns of, simultaneously, several thousands of genes. Thus, they allow  us to study how the genes function and how they perform under different conditions. 
The raw gene expression data consists of an array of quantitative data extracted from one or more chips. For microarrays the data is obtained from scanned images of the chips where each value represents a spot or probe luminosity measured on some unknown scale.  
One of the early observations in gene expression research was that even among chips that were subject to the same processing under
the same conditions by the same lab there could be big differences in the gene expressions that were produced by the experiment.
These differences tend to be systematic and they reflect a lack of a unified chip to chip measurement scale: the measurement scales are homogeneous within a chip but not between chips. 
Technological advances in chip design and automation have mitigated these differences but the improvements have not  succeeded
to eliminate this problem.
In RNAseq similar problems occur in the sense that the scale are not homogeneous. In addition another problem in RNAseq is that often the samples contain many zero values and it is standard to delete those samples with too many zero values and keep the rest for analysis. 

Therefore, in order to transform gene expression data into a format suitable for analysis, we have to preprocess it.  In the preprocessing, the raw data is transformed into a  scale adequate for analysis. In addition, the systematic sources of variation are removed and the outlier observations as well as the gene expressions that are outliers with respect to the given sample should be found.  Due to the particular structure of gene expression data, small samples and high dimensional spaces, standard methods cannot be used. This paper is devoted to two aspects of the preprocessing, doing the so called normalization process and finding the outlier gene expressions. 

Section \ref{Norm} provides the required background  on the  normalization process, which makes use of the component-wise median. We discuss in Section  \ref{nd} the lack of appropriateness of the component-wise median as a robust measure of centrality and introduce an improved normalization process obtained by using the concept of statistical data depth. Gene expression data are known for containing outliers even after normalization that are difficult to detect.  
It is easy to detect severe outliers in individual variables but it is more difficult to detect outliers that may not show up in individual variables but are hidden in the high dimension. 
Thus, although seeking for outlier genes in a given gene expression is a topic that has been broadly studied,  
there has not been put much attention on finding gene expression samples that are outliers. In fact, there are only some graphical techniques based on principal components analysis or spectral maps that could give us an idea about the outliers, 
for instance, the multidimensional scaling graph 
and the biplot of spectral map analysis \cite{CabreraLibro2, Wouters}. None of these two approaches is exact as they are based on a reduction of the  dimensionality and outliers are detected only if they appear in such projections as outliers. Section \ref{om} presents the selection procedure of the outlier gene expressions in a gene expressions' sample. 

Section \ref{realdata} analyzes the below mentioned real data sets using the methodologies presented in Sections \ref{nd} and \ref{om}, for the preprocess of gene expression data. 
The proposed methodologies and employed datasets have been included in an R-package we have named \emph{fdaRNA}. 

\subsection{Background on the Datasets}\label{data}
 The procedures we propose will be applied to different datasets  which are described in what follows. They are the Airway dataset \cite{Himes}, two sets of the Slc17A5 experiment \cite{Raghavan} and \cite{CabreraLibro2} and two known data sets: the called Khan \cite{Khan} and Tissue data sets \cite{CabreraLibro}. 
The Airway dataset is from an RNA-Seq experiment on airway smooth muscle on cell lines that were treated with dexamethasone. The data was extracted from the $airway$ R-package obtained from the Bioconductor repository. It consists of 8 samples, 4 treated and 4 control,  and 29391 RNA's that have more than one nonzero values. About half of the RNA's of the original data had two or more samples with a zero value and were discarded. 
The Slc17A5 data set, which we will refer to as Sialin, compares the gene expression profiles of 6 knocked out mice, mice for which the S1c17A5 gene has been knocked out, versus 6 wild type mice. This is a microarray experiment that was repeated at two time points, one in day first of development during the embryo stage and another at 18 days of age during the adult stage of the mice. 
The  biology suggests that there should be small differences after just a few hours of gestation and larger differences after 18 days. 
Therefore, the  data available for each of the two time points consists of  12 microarrays divided in two groups of six, with 45,101 genes each.
Note the large amount of genes and the small number of microarrays. 
 The Khan data set is a well-known dataset consisting of gene expression measurements, obtained using cDNA microarrays, of four types of pediatric small round blue cell tumors: 
the Ewing family of tumors, Burkitt lymphomas, neuroblastoma and rhabdomyosarcoma. It has 6,567 genes that were filtered in \cite{Khan} to 2,308 genes. We make use of the filtered sample. Particularly, we use the 63 microarrays that form the training set, which contains 23 microarrays from the first class, 8 from the second, 12 from the third and 20 from the last one.
The Tissue data set consists of  the study of the tissue, on mice, using three different treatments. It has 3,487 genes and 41 microarrays. 
Treatment one and two contain each 11 microarrays and treatment three 19 microarrays. 
To summarize the information provided on the datasets, see Table\ref{Tdata}.

\begin{table}[!htb]
\begin{tabular}{r|l}
\hline
Dataset & Experiment on airway smooth muscle on cell lines 
\\ & treated with dexamethasone
\\
Dataset identifier & Airway 
\\
Type of experiment & RNA-Seq
\\
Experimental design& Two groups comparison
\\
Dimension& 8 (4+4) samples and 29391 RNA's 
\\
Filtering procedure& No filtering applied
\\
Dimension after filtering& Does not apply
\\
Raw data & R-package \emph{fdaRNA} and \emph{airway}
\\
Reference & \cite{Himes}
\\
\hline
Dataset & Two sets of the Slc17A5 experiment 
\\
Dataset identifier & Sialin
\\
Type of experiment & microarray
\\
Experimental design& Two groups comparison
\\
Dimension& 2 sets of 12 (6+6) microarrays  with 45,101 genes 
\\
Filtering procedure& No filtering applied
\\
Dimension after filtering& Does not apply
\\
Raw data & R-package \emph{fdaRNA}
\\
Reference& \cite{Raghavan} and \cite{CabreraLibro2} 
\\
\hline
Dataset & Khan training set
\\
Dataset identifier & Khan
\\
Type of experiment & microarray
\\
Experimental design& Four groups comparison
\\
Dimension&  63 (23+8+12+20) microarrays with 6,567 genes
\\
Filtering procedure& quality filtering for a minimal level of expression
\\
Dimensionality after filtering& 63 (12+20+23+8) microarrays with 2,308 genes
\\
Raw data & R-package \emph{fdaRNA}
\\
Reference& \cite{Khan} 
\\
\hline
Dataset & Study of the tissue, on mice, using three different treatments 
\\
Dataset identifier & Tissue
\\
Type of experiment & microarray
\\
Experimental design& Three groups comparison
\\
Dimension& 41 (11+11+19) microarrays with 45,101 genes 
\\
Filtering procedure& Fold ratio over 2
\\
Dimensionality after filtering& 41 (11+11+19) microarrays with 3,487 genes 
\\
Raw data & R-package \emph{fdaRNA}
\\
Reference&  \cite{CabreraLibro}
\\
\hline
\end{tabular}
\caption{Table describing the features of the datasets used in the study: dataset identifier in the manuscript, type of experiment (such RNA-seq or microarray), a brief description of the experimental design, dimensionality, filtering procedures that were applied, dimensionality after filtering, R-package for  the raw data and the reference.}\label{Tdata}
\end{table}

\begin{table}[!htb]
\begin{center}
\begin{tabular}{cccccc}
\multicolumn{6}{c}{Airway} 
\\ 
\hline
SRR1039508 &SRR1039509 & SRR1039512 &SRR1039513& SRR1039516 &SRR1039517
\\
1&2&3&4&5&6
\\
 SRR1039520 &SRR1039521&&&&
\\ 
7&8&&&&
\\
\hline
\\
\multicolumn{6}{c}{Sialin after 6 hours}
\\
\hline
X2760.CEL  & X2761.CEL  &  X2762.CEL  &  X2763.CEL  &  X2764.CEL   & X2765.CEL  
 \\ 
 1&2&3&4&5&6
 \\
X2766.CEL   & X2767.CEL  &  X2768.CEL  &  X2769.CEL   & X2770.CEL   & X2771.CEL
 \\
7 &8&9&10&11&12
\\
\hline
\\
\multicolumn{6}{c}{Sialin after 18 days}
\\
\hline
X2720.CEL   & X2721.CEL   & X2722.CEL   & X2723.CEL   & X2724.CEL   & X2725.CEL  
 \\ 
 1&2&3&4&5&6
 \\
X2726.CEL  &  X2727.CEL   & X2728.CEL   & X2729.CEL   & X2730.CEL  &  X2731.CEL
 \\
7 &8&9&10&11&12
\\
\hline
\end{tabular}
\end{center}
\caption{RNA-seq or microarray  identifier used in the R-package \emph{fdaRNA} for  the Airway and Sialin datasets and associated identification number  used in the tables and figures of this paper.}\label{names1}
\end{table}

\begin{table}[!htb]
\begin{center}
\begin{tabular}{cccccc}
\multicolumn{6}{c}{Khan}
\\
\hline
TRAIN1.EW & TRAIN2.EW & TRAIN3.EW & TRAIN4.EW & TRAIN5.EW & TRAIN6.EW
 \\ 
 1&2&3&4&5&6
 \\
 TRAIN7.EW & TRAIN8.EW & TRAIN9.EW & TRAIN10.EW & TRAIN11.EW & TRAIN12.EW 
 \\
7 &8&9&10&11&12
\\ 
TRAIN13.EW & TRAIN14.EW & TRAIN15.EW & TRAIN16.EW & TRAIN17.EW & TRAIN18.EW 
 \\
 13&14&15&16&17&18
\\ 
TRAIN19.EW & TRAIN20.EW & TRAIN21.EW & TRAIN22.EW & TRAIN23.EW & TRAIN24.BL 
 \\
 19&20&21&22&23&24
\\ 
TRAIN25.BL & TRAIN26.BL & TRAIN27.BL & TRAIN28.BL & TRAIN29.BL & TRAIN30.BL
 \\
 25&26&27&28&29&30 
\\ 
TRAIN31.BL & TRAIN32.NB & TRAIN33.NB & TRAIN34.NB & TRAIN35.NB & TRAIN36.NB 
 \\
 31&32&33&34&35&36
\\
 TRAIN37.NB & TRAIN38.NB & TRAIN39.NB & TRAIN40.NB & TRAIN41.NB & TRAIN42.NB 
  \\
 37&38&39&40&41&42
 \\
  TRAIN43.NB & TRAIN44.RM & TRAIN45.RM & TRAIN46.RM & TRAIN47.RM & TRAIN48.RM 
   \\
 43&44&45&46&47&48
  \\
   TRAIN49.RM & TRAIN50.RM & TRAIN51.RM & TRAIN52.RM & TRAIN53.RM & TRAIN54.RM 
    \\
49 &50&51&52&53&54
   \\
    TRAIN55.RM & TRAIN56.RM & TRAIN57.RM & TRAIN58.RM & TRAIN59.RM & TRAIN60.RM
     \\
 55&56&57&58&59&60
    \\
     TRAIN61.RM & TRAIN62.RM & TRAIN63.RM &&&\\
 61 &62&63&&&
 \\
\hline
\\
\multicolumn{6}{c}{Tissue}
\\
\hline
X1 & X2  &  X3  &  X4   & X5  &  X6  
 \\ 
 1&2&3&4&5&6
 \\
  X7  &  X8&    X9 &  X10&   X11 &  X12 
 \\
7 &8&9&10&11&12
\\ 
   X13  & X14& X15  & X16  & X17  & X18  
 \\
 13&14&15&16&17&18
\\ 
 X19  & X20   &X21 &  X22  & X23  & X24 
 \\
 19&20&21&22&23&24
\\ 
  X25  & X26   &X27 & X28  & X29  &X30 
 \\
 25&26&27&28&29&30 
\\ 
  X31 &  X32 &  X33  & X34  & X35  & X36  
 \\
 31&32&33&34&35&36
\\
 X37  &X38  & X39  & X40  & X41&
 
  \\
 37&38&39&40&41&
\\
\hline
\end{tabular}
\end{center}
\caption{Microarray  identifier used in the R-package \emph{fdaRNA} for  the Khan and Tissue datasets and associated identification number  used in the tables and figures of this paper.}\label{names2}
\end{table}

\subsection{Background on Normalization}\label{Norm}

Normalization was proposed as a way to correct the measurement scale differences of gene expressions.
It can be regarded as a sort of calibration process that improves the comparability among gene expressions treated alike.
Initially, biologists applied  normalization using simple methods. They were linear transformations that corrected the systematic measurement differences by equating either the mean, the
sum of each sample, the component-wise median or the component-wise third quartile.  The common feature of these normalization schemes was
that they assume that the spot intensities on every pair of arrays being normalized are linearly related with no intercept,
so that the lack of comparability can be corrected by adjusting every single spot intensity on any gene expression by the same
amount, called the normalizing factor, regardless of its intensity level.
However, the relationship between the spot intensities of two different gene expressions is usually nonlinear. Thus, an intensity dependent normalization method, i.e. a normalization scheme in which the normalization factor is a function of intensity level, should be preferable to a linear normalization method.
In intensity dependent normalization, the transformed spot intensity data is normalized using a nonlinear normalization function. 
Nonlinear  and quantile normalizations were 
introduced in  \cite{Cabrera1} and \cite{Cabrera1a}, 
under the label of quantile and non-linear standardization, 
and in \cite{Bolstad}, 
who coined the term ``normalization''.
 Recent publications that regard normalization of gene expression data include  \cite{Wu419}  that presents an R-package  for the evaluation of normalization methodologies, \cite{FUNDEL082008, Ab102018} that compares different normalization methods
and \cite{CabreraLibro2}  for an analysis of the different normalization approaches. Further publications are \cite{Li, Sh, Sc}.

 The different ways to define non-linear normalization have the following  commonalities.
Let $X=\{ x_{ij}\}$ be a data matrix of gene expressions with i=1,...,G genes and j=1,..,n samples. We denote by $X_{\cdot j}$ the $j-$th sample. Suppose that each column of X is subject
 to slightly different measurement scales. This assumes that there is a data matrix $Z=\{z_{ij}\}$ such that $x_{ij} = h_j(z_{ij}),$ with the $h_j 's$  nonlinear monotonic transformations $h_j : \mathbb{R}\rightarrow \mathbb{R}$. 
 There are two issues with the estimation of $h_j$'s, the first one is that the $h_j$ functions are not identifiable up to a monotonic transformation and the second is the nature of the error term, which one of $X$ or $Z$ is observed with error.
The identifiability issue on $h_j$ is usually overcomed by establishing an ideal scale or mock array that represents a universal standard.  The process is as follows.
\begin{enumerate}
\item
Do a basic linear transformation to the columns of X that equates the component-wise median or component-wise 75\% percentiles, for example.
\item
Calculate the reference array $\mu$.  For instance, \\  
$\mu= \{ \mu_i =\mbox{median}(x_{i1}, \dots , x_{in}) : i=1, \dots, G\} $.
\item
 Back transform each of the columns of $X$ to the scale defined by $\mu$.
\end{enumerate}

\subsubsection{\emph{Quantile Normalization}.}\label{normalization}

Quantile normalization is applied in order to make the distributions of the gene expression $\{x_{ij}\}$ as similar as possible across the columns of $X$ by making it similar to the distribution of gene expressions of the component-wise median array.  Usually all the component-wise quantiles are equated although one may chose to equate only a subset of the quantiles.
The following algorithm  is used to equate a subset of quantiles, say quantiles ($Q^j_0, \dots , Q^j_L$) of the $j-$th vector and the quantiles ($Q^M_0, \dots , Q^M_L$) of the component-wise median vector \cite{Cabrera1}.
\begin{enumerate}
\item
Let $X^*$ be obtained from  $X$ by sorting the columns in ascending order.
\item \label{dos}
Let $X^M=(x^M_1, \dots, x^M_G)$ be  the vector of row medians of  $X^*$.
\item
For any  $x_{ij}$, $i=1, \dots , G, j=1, \dots ,n$, find the interval,
$ [Q^j_{k}$,$ Q^j_{k+1}]$, to which it belongs to.
\item
Obtain the normalized value of $x_{ij}$, by linearly interpolating between the pair of points: $(Q^M_k$, $Q^j_{k}$) and ($Q^M_{k+1}$, $Q^j_{k+1}$).
\end{enumerate}
The algorithm  for equating all quantiles of the columns of $X$ to the quantiles of the component-wise median array is simpler  \cite{Cabrera1}: It suffices to substitute 3.~and 4.~above by:
\begin{enumerate}
\setcounter{enumi}{2}
\item
For any value $x_{ij}$, $i=1, \dots , G, j=1, \dots ,n$, find its rank $r$ with respect to the $j-$th column. 
\item The normalized value of $x_{ij}$ is $x^M_r.$
\end{enumerate}

\section{Methods}
\subsection{Normalization using statistical functional depth}\label{nd}

\begin{figure}[!htbp]
 \includegraphics[width=.5\linewidth]{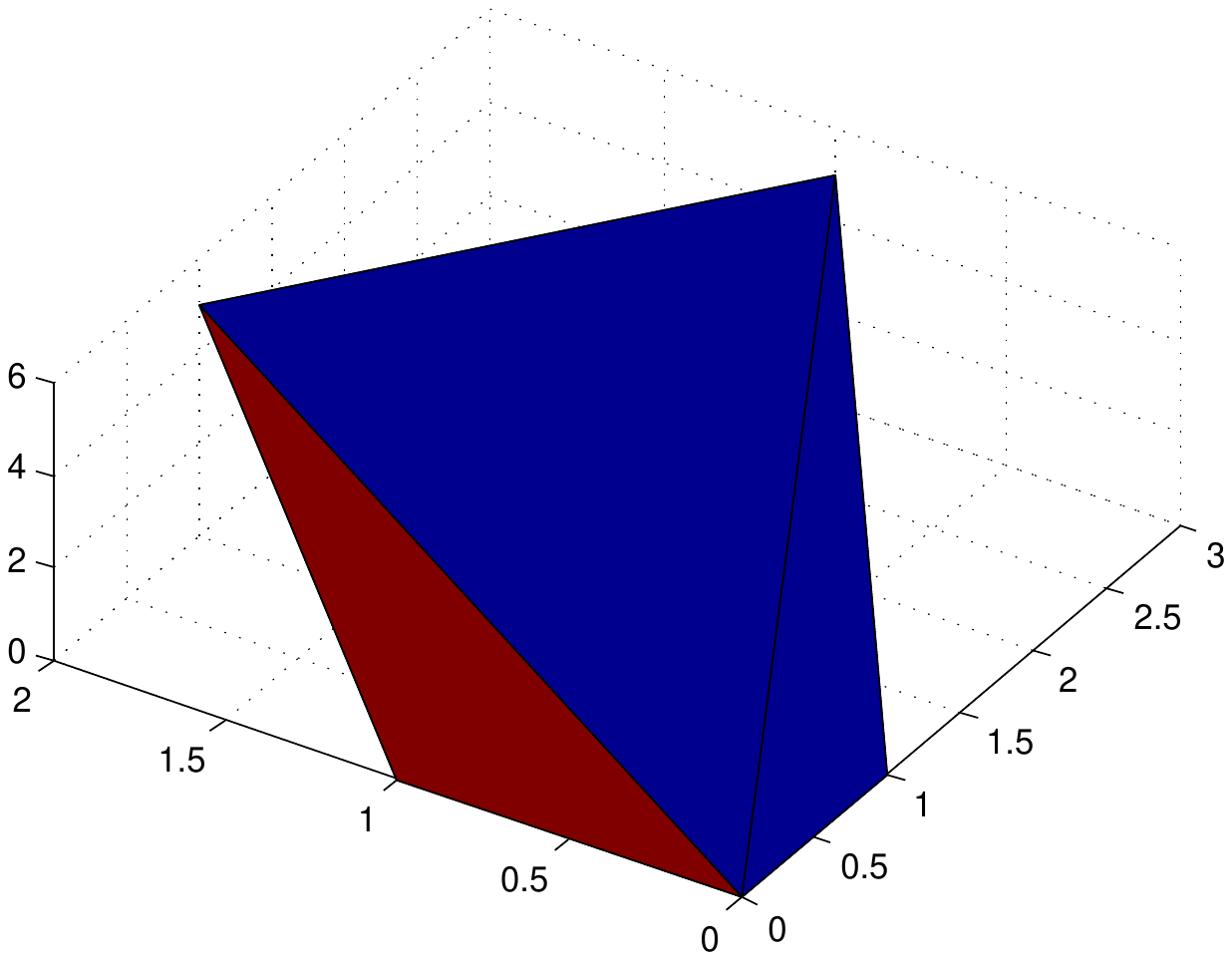}
 \includegraphics[width=.5\linewidth]{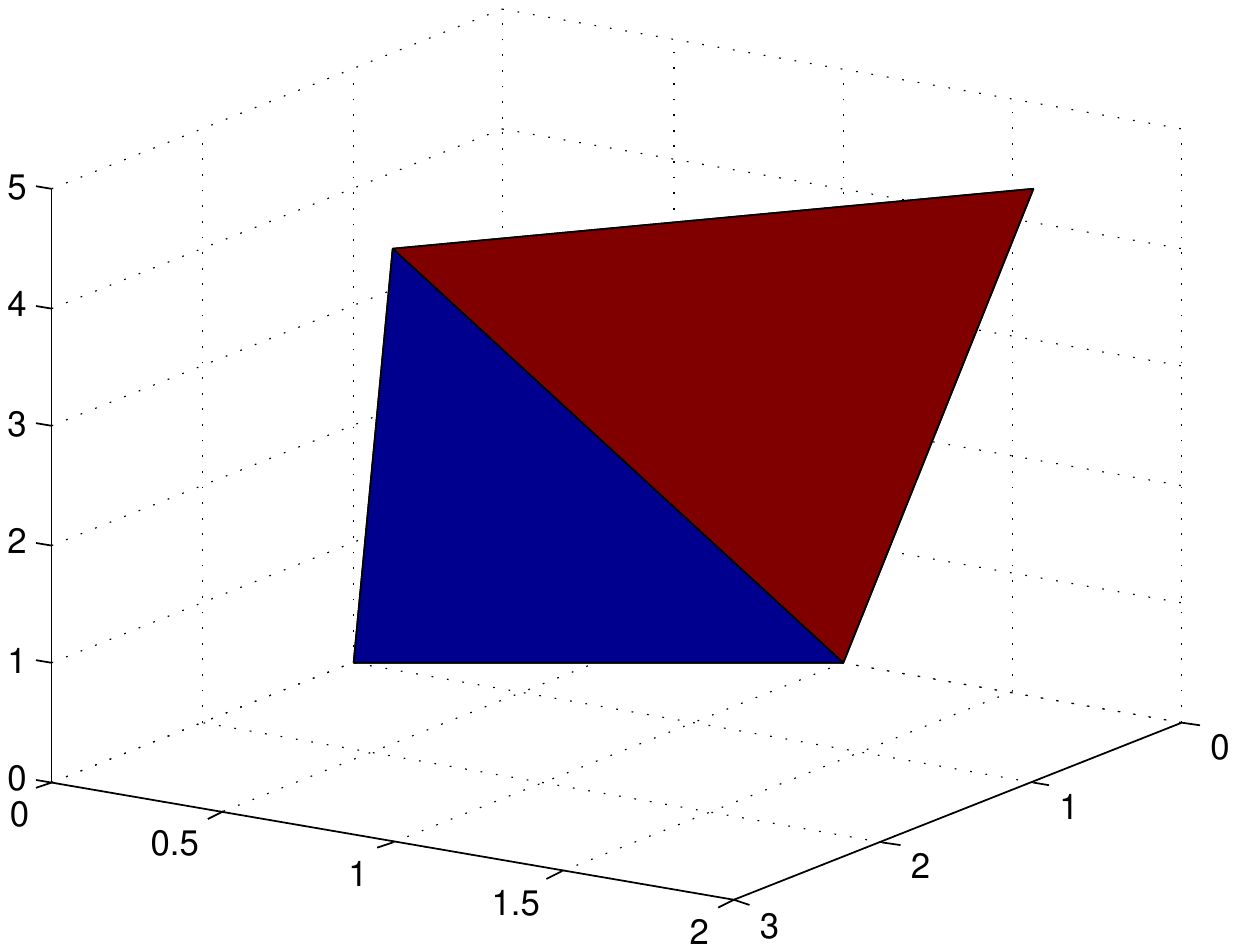}
 \vspace{-3.5cm}
\caption{Convex hull of the vectors $(0,1,0),$ $(0,0,0),$ $(1,0,0),$ $(1,2,5)$ and $(3,1,5)$ in two different perspectives. The coordinate wise median of these vectors does not belong to the convex hull. The x-axis varies in the 0-3 range and  the y-axis in 0-2 range. The z-axis varies in the 0-6 range In the left plot and in the 0-5 range in the right plot. The colors have no meaning; their aim is to make a distinction  among the different faces of the convex hull.} 
\label{convhull}
\end{figure}

The main issue with quantile normalization is to determine $X^M.$  
It is well know in multivariate analysis that the coordinate-wise median is not an 
adequate measure of centrality. To order a data set 
in $\Rea,$ it is reasonable to
follow the decreasing order of the
absolute values of the difference between the percentiles of the points in the set and the percentile 50. Thus, the data set central points are the median(s) of the dataset.  In contraposition, the component-wise median of a sample of gene expressions  is not necessarily  similar to any of the gene expressions in the sample.
This is because, if our data set is in $\Rea^p,$ $p>1,$ the median has to be calculated coordinate by coordinate. It is well known that the component-wise median may not always
 give an idea of centrality, 
and in some cases even not fall inside the convex hull of the data.
The following two examples illustrate these cases.
Let us take  in $\Rea^2$
the vertices of an equilateral triangle and its center of mass, all with equal probability. By symmetry, 
the deepest point should be unique and coincide with the center of mass. However, that is not the case of  the component-wise median, an area that depends on the selected coordinate axes.
For the second example, in Figure \ref{convhull} it is displayed the convex hull of the vectors $(0,1,0),$ $(0,0,0),$ $(1,0,0),$ $(1,2,5)$ and $(3,1,5)$ in two different perspectives. The component-wise median of these vectors is $(1,1,0)$ and, as it can be appreciate from the figure, if does not fall inside the convex hull of the data.

We present an alternative normalization methodology that uses the notion of statistical depth. \cite{Tukey}  
introduced this notion as a way to emulate the behavior  of the one-dimensional quantiles, and median, to give a concept of order valid for spaces whose 
dimension is higher than one. Informally, the data is ordered such that, if a
datum is moved toward the center of the data cloud, its statistical depth
increases, and, if it is moved toward the outside, its
statistical depth decreases.
Thus, finding the deepest point in a set is the same as finding  the point that is most inside the points cloud or, roughly speaking, the one with most points around it.
 Tukey's idea was for  statistical depth to be a robust measure that would yield a definition
of ``multivariate median'' (deepest element(s)) that was robust, as an alternative to earlier 
ideas like convex peeling and erosion peeling that lacked
resistance to pattern of outliers and therefore had a 
breakdown point of zero.
This work was followed by  \cite{Donoho92} and \cite{Stahel}  
who simultaneously introduced the methodology of projection depth and they achieve a ``multivariate median'' with 50\% breakdown point.
Later, some requirements that every multidimensional statistical depth 
should fulfill were established in \cite{Zuo00}. Several definitions of multidimensional statistical depth 
have been proposed. Some of the best-known are Tukey's depth (or  half-space depth), Oja's depth and simplicial depth
(see, for instance,  \cite{LiuGriego}).

\begin{figure}[!htbp]
 \includegraphics[width=.5\linewidth]{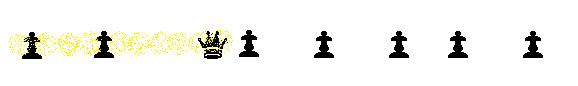}
 \caption{Representation of a data sample in dimension one by a king and pawns. The one-dimensional statistical depth value of the king with respect to the sample, 3/8, is illustrated with yellow spots.} 
\label{Peones1}
\end{figure}

\begin{figure}[!htbp]
 \includegraphics[width=.5\linewidth]{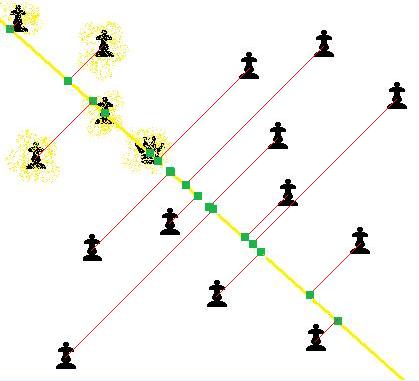}
 \caption{Representation of a data sample in dimension two by a king and pawns, a thick yellow line in which to project the pawns and the projected pawns (green diamonds at the end of the slim red segments). The statistical depth value, restricted to the thick yellow line,  of the king with respect to the sample is illustrated with yellow spots. It takes value 5/16.} 
\label{Peones2}
\end{figure}

Given a sample of data, to exemplify the Tukey's depth of an element of the sample with respect to the sample, we have represented in Figures \ref{Peones1} and \ref{Peones2} a set of data points by pawns and a king. We aim to compute the depth of the data point represented by the king. Such depth is computed with respect to the data points represented by the pawns and the king itself. Figure \ref{Peones1} exemplifies Tukey's depth in dimension one and Figure \ref{Peones2} in dimension two. Let us first focus on Figure \ref{Peones1}. Computing Tukey's depth  in dimension one is equivalent to first calculate the minimum between  the number of pieces that have to be captured to capture the king if we approach it by the right and if we do so by the left. Then, this number is divided by the total number of pieces. In the plot, we obtain that this number is the minimum between three (from the left) and six (from the right). Then, the depth of the king with respect to the plotted pieces is 3/8. 
In higher dimensions, $p>1$, Tukey's depth is computed by projecting the data in all possible vectors, of norm one, of the sphere of dimension $p-1$ (it reduces to a finite amount of vectors when computed with respect to a sample) and computing the minimum of all these resulting one-dimensional depth values. Thus, in  Figure  \ref{Peones2} it is plotted a yellow line determined by one of the possible vectors and by the data point represented by the king. The procedure consists on projecting each pawn in the yellow line (the projected pawns are the green diamonds at the  end of each red segment) and then computing the one-dimensional Tukey's depth. In the plotted line, the one-dimensional Tukey's depth of the king with respect to the plotted pieces is 5/16.
A further example is the equilateral triangle and its center of mass commented at the top of this section. In that case, the center of mass obtains uniquely the highest Tukey's depth.

The  objective of gene expression normalization is to equate the scales at which the various gene expressions have been measured. Each scale is a continuous curve that is approximated by a non-decreasing vector that is obtained by ordering the spot intensities of a gene expression.
Thus, we propose to choose as $X^M$ the deepest gene expression among  $X^*_{\cdot 1}, \ldots,X^*_{\cdot n},$ which will preserve the idea of centrality and the simplicity of the procedure while lacking  of the cited problematic of the component-wise median. Then, our proposal modifies the normalization procedure of Subsection \ref{normalization} substituting item \ref{dos}. by:

\ref{dos}.*  Let $X^M$ be the deepest element of the set $X^*_{\cdot 1}, \ldots,X^*_{\cdot n}.$ \\ The dimensionality of such vectors could easily reach over half a million because a typical gene expression has 45K-50K genes and for each gene we may observe eleven probes. Thus, classical multidimensional statistical depth functions are not applicable from a practical point of view because of their computational cost \cite{Mosler}. From a theoretical point of view, in \cite{Indios} it is studied the non appropriateness of using multidimensional statistical depth functions either, due to the  high dependency structure among closed by observations  that occur in this type of data.
Using parallel coordinates \cite{Inselberg}, $X^*_{\cdot 1}, \ldots,X^*_{\cdot n} $ can be viewed as the partial observations of  non-decreasing curves.
 Consequently, it is appropriate to use statistical functional depths  (see \cite{Nieto-Battey} for a review on the matter), which are not time consuming and lack of the theoretical issues expressed in \cite{Indios}.  

Throughout this article, we use the definition of statistical functional depth introduced in \cite{LibroSantander}. This definition is based on a distance. For its simplicity and interpretability, we propose to use throughout this paper the $L_2$ space and distance, which coincides with the euclidean because of the use of parallel coordinates. Therefore, what makes this statistical functional depth behave well is the fact that the regions in which the curves are close have less weight in computing the distances among curves than other regions. 
In addition, this definition of statistical functional depth   is robust in the sense that it does not matter how far away is a curve, or a region of the curve, from the curves cloud as it is in the same layer as if it were closer.

The formal definition  follows. Given a  set of functions, $X:=\{X_{\cdot 1}, \ldots, X_{\cdot n}\}\subset L_2,$ we aim to compute the depth of $X_{\cdot j}$ with respect to $X$ for each $j=1, \ldots, n.$
This statistical functional depth computes the statistical functional deepness of each of these functions with respect to the set by constructing a sequence of layer sets, $\frak{L}_k,$ and associated borders,  $\frak{B}_k.$
First, it is defined a layer and associated border formed by the most distant functions (or curves).
$$\frak{L}_1:=\{X_{\cdot}\in L_2:d(X_{\cdot},X_{\cdot 1_i})\geq d(X_{\cdot 1_1},X_{\cdot 1_2}) \mbox{ for an } i=1,2; X_{\cdot 1_1},X_{\cdot 1_2}\in\frak{B}_1\},$$ with border
$\frak{B}_1:=\{\arg \max d(X_{\cdot i},X_{\cdot j}): X_{\cdot i},X_{\cdot j}\in X\}.$
Then, subsequent layers are defined inside the previous ones by the subsequent most distant curves. 
$$\frak{L}_k:=\{X_{\cdot}\in L_2-\cup_{l=1}^{k-1}\frak{L}_l:d(X_{\cdot},X_{\cdot k_i})\geq d(X_{\cdot k_1},X_{\cdot k_2}), i=1  \mbox{ or } 2; X_{\cdot k_1},X_{\cdot k_2}\in\frak{B}_k\},$$
with border $\frak{B}_k:=\{\arg \max d(X_{\cdot i},X_{\cdot j}): X_{\cdot i},X_{\cdot j}\in X-\cup_{l=1}^{k-1}\frak{B}_l\}, \mbox{ for }k=2,...,k_0.$ 
$k_0$ is such that $X-\cup_{l=1}^{k_0}\frak{B}_l$ contains only one curve or is equal to $\frak{B}_{k_0+1}.$  Thus, the statistical functional deepest curve/s among $X_{\cdot 1}, \ldots, X_{\cdot n}$ is/are 
in $\frak{B}_{k_0+1}$ 
and $\frak{B}_{k}$ is empty for $k>k_0+1.$
While the statistical functional  depth of any general element of $L_2$ requires the use of these layers, it suffices to make use of the borders to compute the depth of the sample curves. Thus, according to \cite{LibroSantander}, for any $j=1, \ldots, n,$ the statistical functional depth of $X_{\cdot j}$ with respect to $X$ is 
$$D(X_{\cdot j}, X):=\{k\in\{1, \ldots, k_0+1\}: X_{\cdot j}\in\frak{B}_k\}/n
;$$ 
the index of the border to which $X_{\cdot j}$ belongs divided by the sample size.
Thus, the statistical functional depth of a sample curve is given by the border to which the curve belongs.

\begin{figure}[!htbp]
  \includegraphics[width=.5\linewidth]{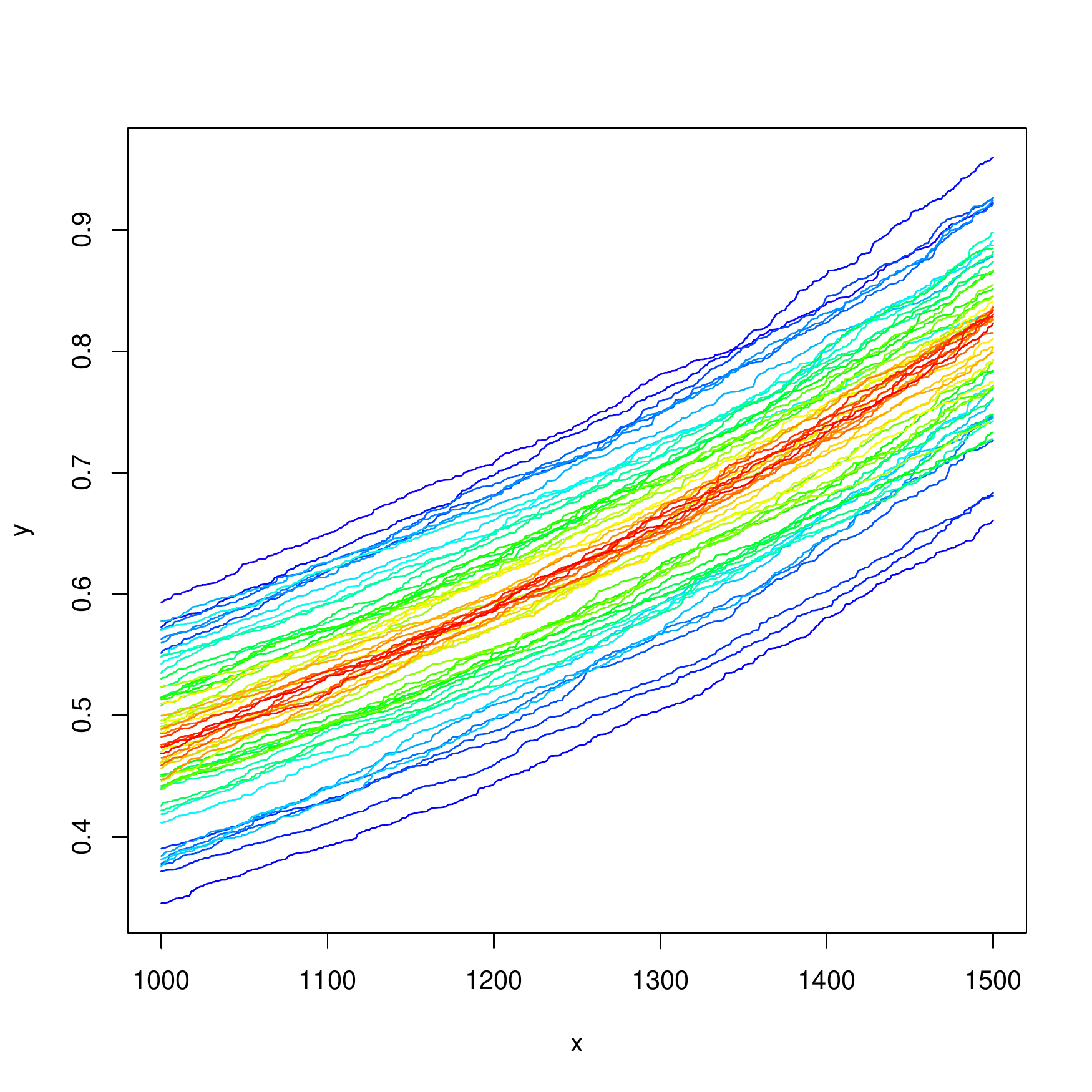}
 \caption{Illustration of the statistical functional depth used in the paper.
 For that, a zoom, of the central genes, for the ordered genes in the process of normalization of the Khan dataset has been displayed.  The colors represented the statistical functional depth of the ordered microarrays, from blue for low statistical  functional depth to red for high statistical functional depth, through green and yellow. It can be appreciated how the border sets are nested in. 
This illustration is complemented by a movie in Figure 1 of the Supplementary material.} 
\label{zoom}
\end{figure}

To illustrate this statistical functional depth we have included in Figure \ref{zoom} a zoom, of the central genes, for the ordered genes in the process of normalization of the Khan dataset. With the colors, it is represented the statistical functional depth of the microarray curves, from blue for low statistical  functional depth to red for high statistical functional depth, through green and yellow. Thus, it can be appreciated how the border sets are nested in. For a better appreciation, we have included as  supplementary material, a movie where it is appreciated the construction of the borders involved in constructing this plot.

More involved instances of statistical functional depth with these characteristics can be sought  (see \cite{Nieto-Battey}). However, the intention here is to propose a straightforward, and easy, to understand and apply methodology that does not require of expert statisticians for its implementation and analysis.

\begin{Nota}
In order to make the component-wise method more usable, the data must be preprocessed by
individual linear transformations of each gene expression vector to make them as similar as possible, otherwise the coordinate-wise median may yield unreasonable results. If we normalize to the deepest array then this problem may not occur and the linear preprocessing is not necessary. However, in practice, the linear preprocessing may yield a different and perhaps more suitable deepest array, so we believe that is not a bad idea.
\end{Nota}

\subsection{Outlier gene expressions}\label{om}

Here we develop an analytical technique to detect outliers specifically design to this kind of data. It is quite simple and so, pleasant for either statisticians and biologist, who has to deal with gene expression data. However, it can be generally used. This technique is based on statistical functional depth and the well-known Tukey's method for outlier detection in $\Rea$ \cite{tukeyLibro}. 
According to Tukey, 
$x\in\Rea$ or $y\in\Rea$ is a potential outlier if \begin{eqnarray}\label{mild}x > q_3 + g \cdot(q_3 - q_1)  \mbox{  or  } y < q_1 -g\cdot (q_3 - q_1),\end{eqnarray} 
 where $q_1$ denotes the lower quartile and $q_3$ denotes the upper quartile of a given sample or distribution. Tukey proposed to take $g$ equal to 1.5 for mild outliers and to 3 for severe outliers. However, later authors propose to take $g$ depending on characteristics of the data such as sample size, see for instance \cite{gather}. 
 By subtracting  the two inequalities in (\ref{mild}),  the statement is rewritten as   \begin{eqnarray}\label{mild2}x-y > (2g+1)\cdot (q_3 - q_1).\end{eqnarray}

Our methodology orders the data in pairs from the statistical least deep pairs to the statistical deepest. Let  $X:=\{X_{\cdot 1}, \ldots, X_{\cdot n}\}$ be a data matrix of gene expressions, with $\frak{B}_k=\{ X_{\cdot k_1},X_{\cdot k_2}\}$ the k-th least deep pair of gene expressions with respect to $X,$ for $k=2,...,k_0.$ Thus, we propose to generalize Inequality \ref{mild2} from $\Rea$ to any general space with a metric $d(\cdot,\cdot),$ by considering the pair $X_{\cdot k_1},X_{\cdot k_2}$ as a potential outlier if 
 \begin{eqnarray}\label{mild3}d(X_{\cdot k_1},X_{\cdot k_2}) > G\cdot d(Q_3 ,Q_1).\end{eqnarray}
 As before, for simplicity, we consider the $L_2$ (euclidean) distance. For the interquartile range, $d(Q_3 ,Q_1),$ we estimate it by computing the 
 $$\mbox{median}\{d(X_{\cdot k_1},X_{\cdot k_2}) : X_{\cdot k_1},X_{\cdot k_2}\in\frak{B}_k,   k\leq k_0+1 \}.$$ 
 If $n$ is odd, $d(X_{\cdot {k_0+1}_1},X_{\cdot {k_0+1}_2})=0.$
 For an example, let us assume we have the data sample $$1.3, 2.1, 2.8,2.9,3.2,3.9,4.1,4.8,4.9,5.3\in\Rea$$  of size 10. Then, we have to compute the $\mbox{median}\{4,2.8,2,1.2,0.7\}$ that is 2, which is the value we would have obtained if computing directly the interquartile range in $\Rea$ of the given sample. 
To estimate $G,$ we perform a Monte Carlo study in which we draw a multivariate normal distribution with the sample size, dimension and covariance structure of the given data. The latter can be estimated robustly. Here, to estimate $G,$  it is estimated the quantile such that a $.01$ percent of the columns  are flagged as outliers. The process is repeated a number of times, 100 times for the results in Section \ref{realdata}, and the median of those quantiles is the estimated value of $G$

If a pair  $X_{\cdot k_1},X_{\cdot k_2}$ satisfies (\ref{mild3}), it is in general because one of the elements of the pair is outlying. Thus, we consider as potential outlier the element of the pair that is farther away from the statistical deepest gene expression of $X,$ for instance in $L_2$ (euclidean) distance. However, a more involve procedure can be undertaken, if the practitioner consider it necessary, in which both elements of the pair would be potential outliers. 
Note that our approach is different from the one of the functional boxplot, see \cite{gentonMC} and the references therein.

\section{Result}
\label{realdata}

\begin{figure}[!htbp]
 \includegraphics[width=.3\linewidth]{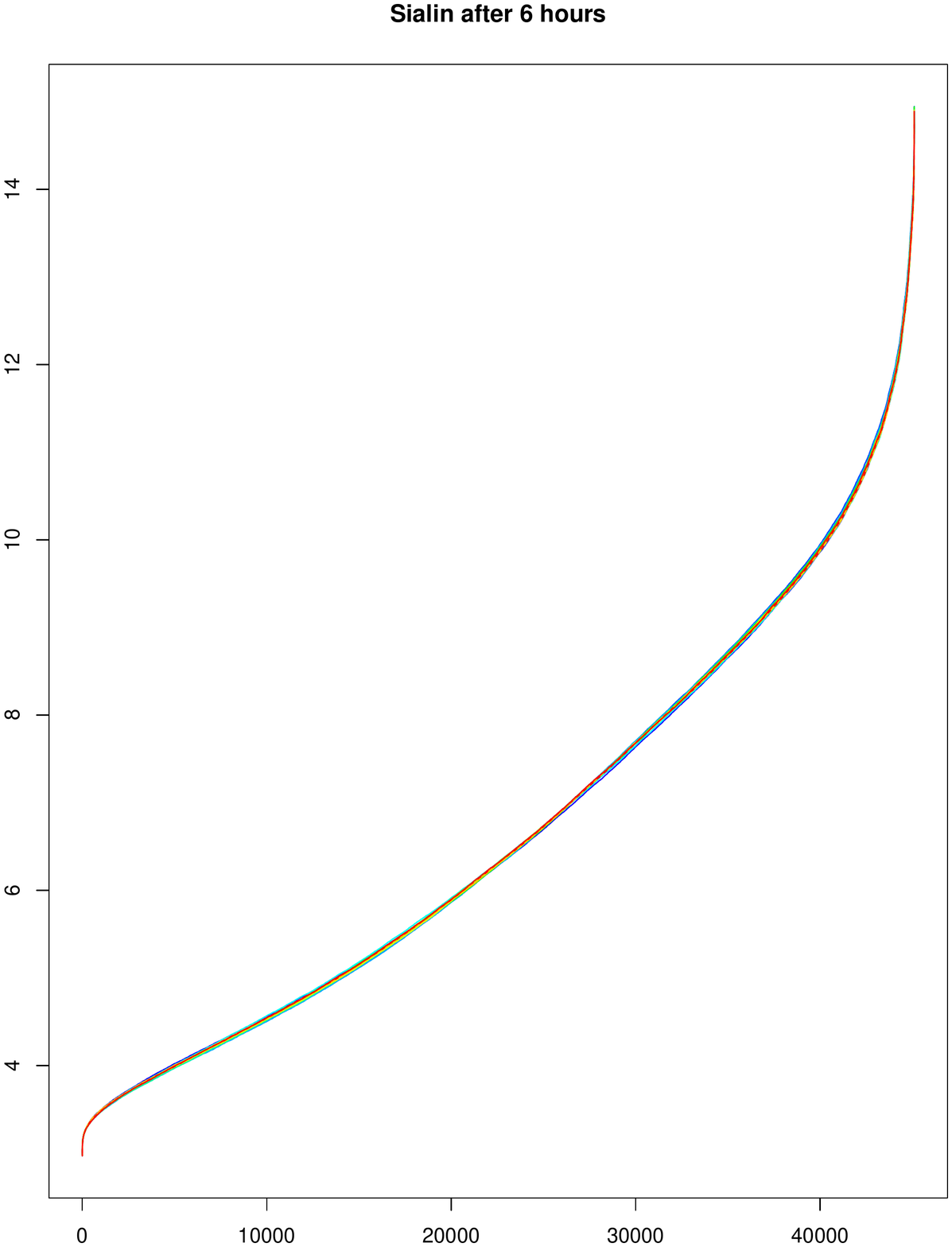}
     \includegraphics[width=.3\linewidth]{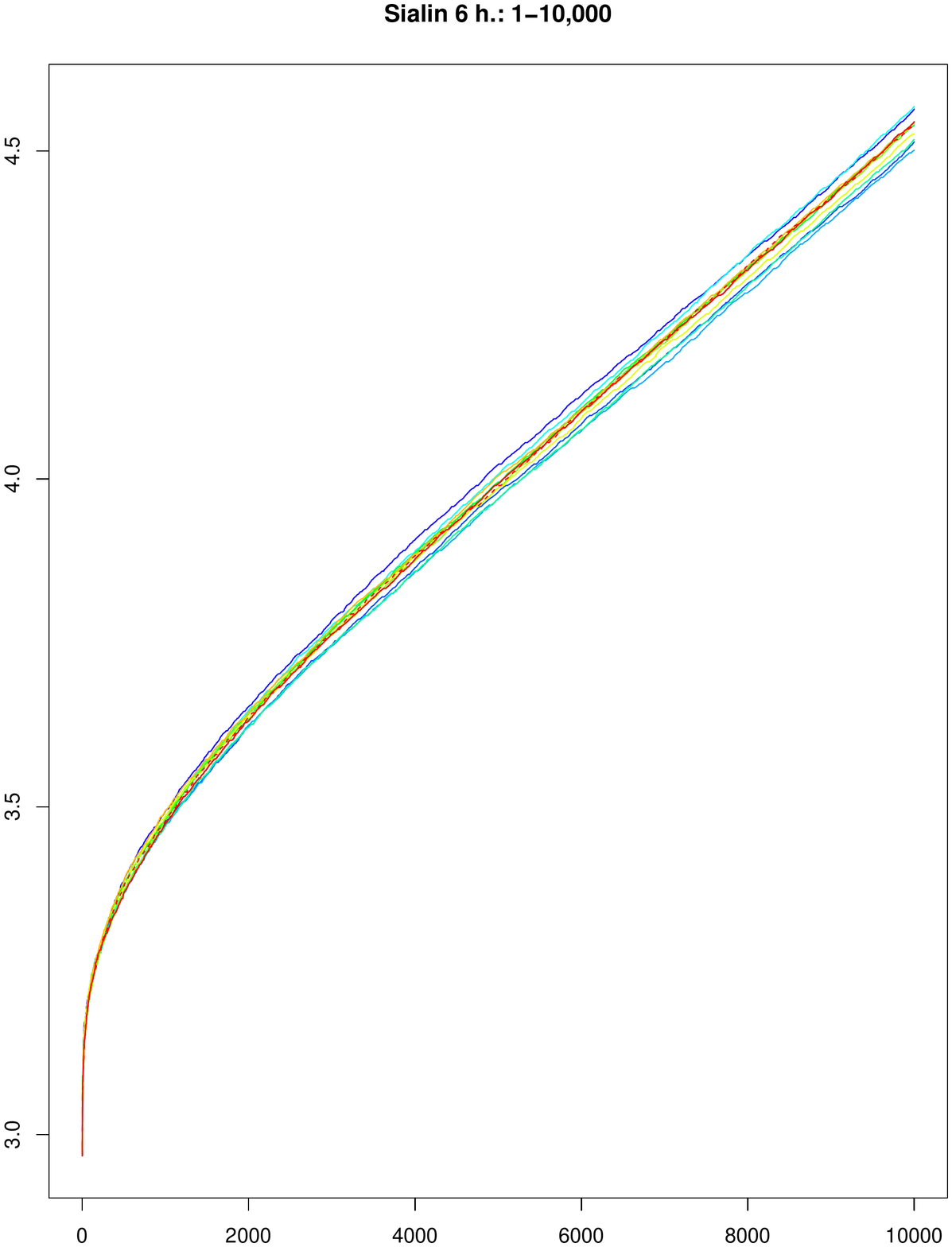}
      \includegraphics[width=.3\linewidth]{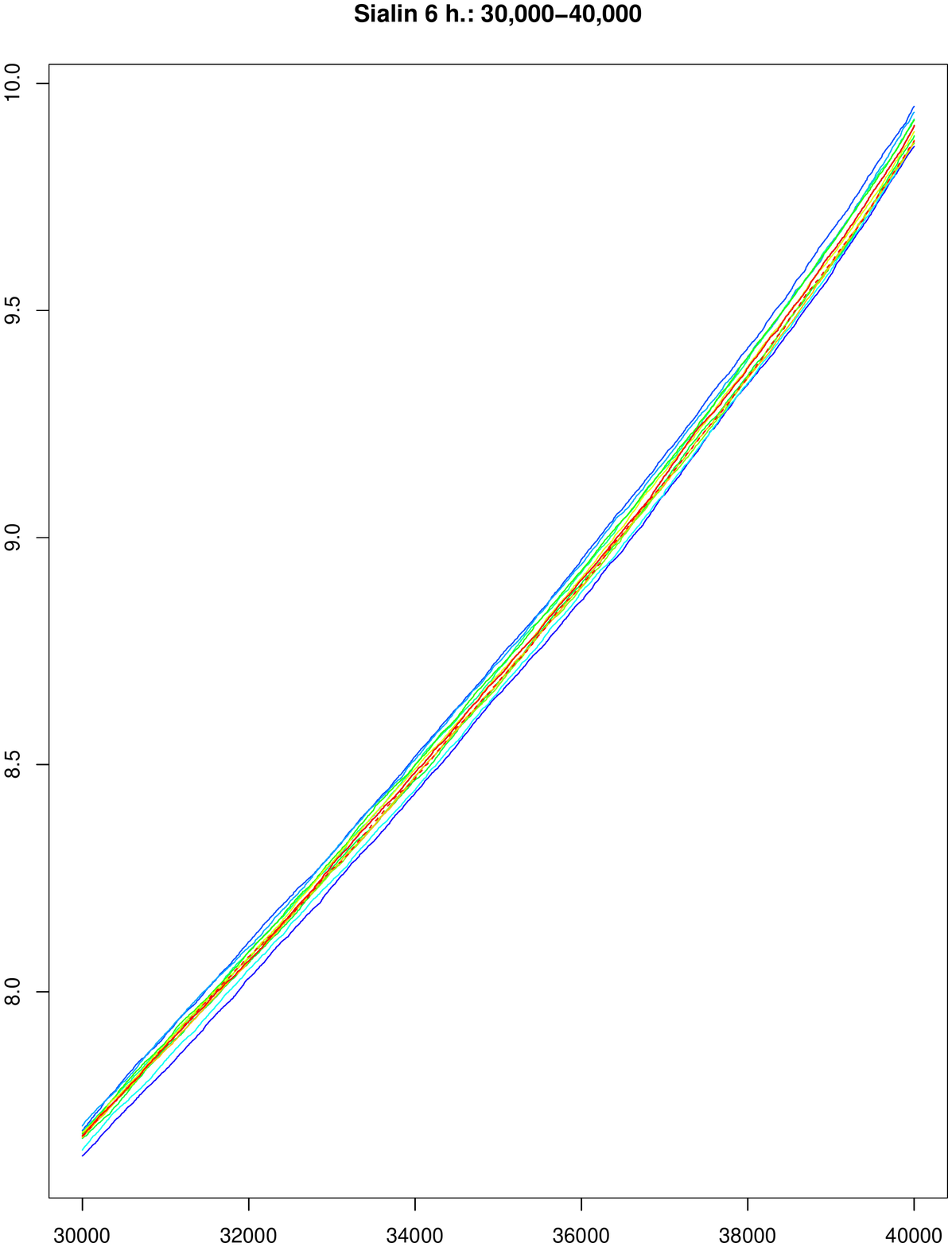}
 \includegraphics[width=.3\linewidth]{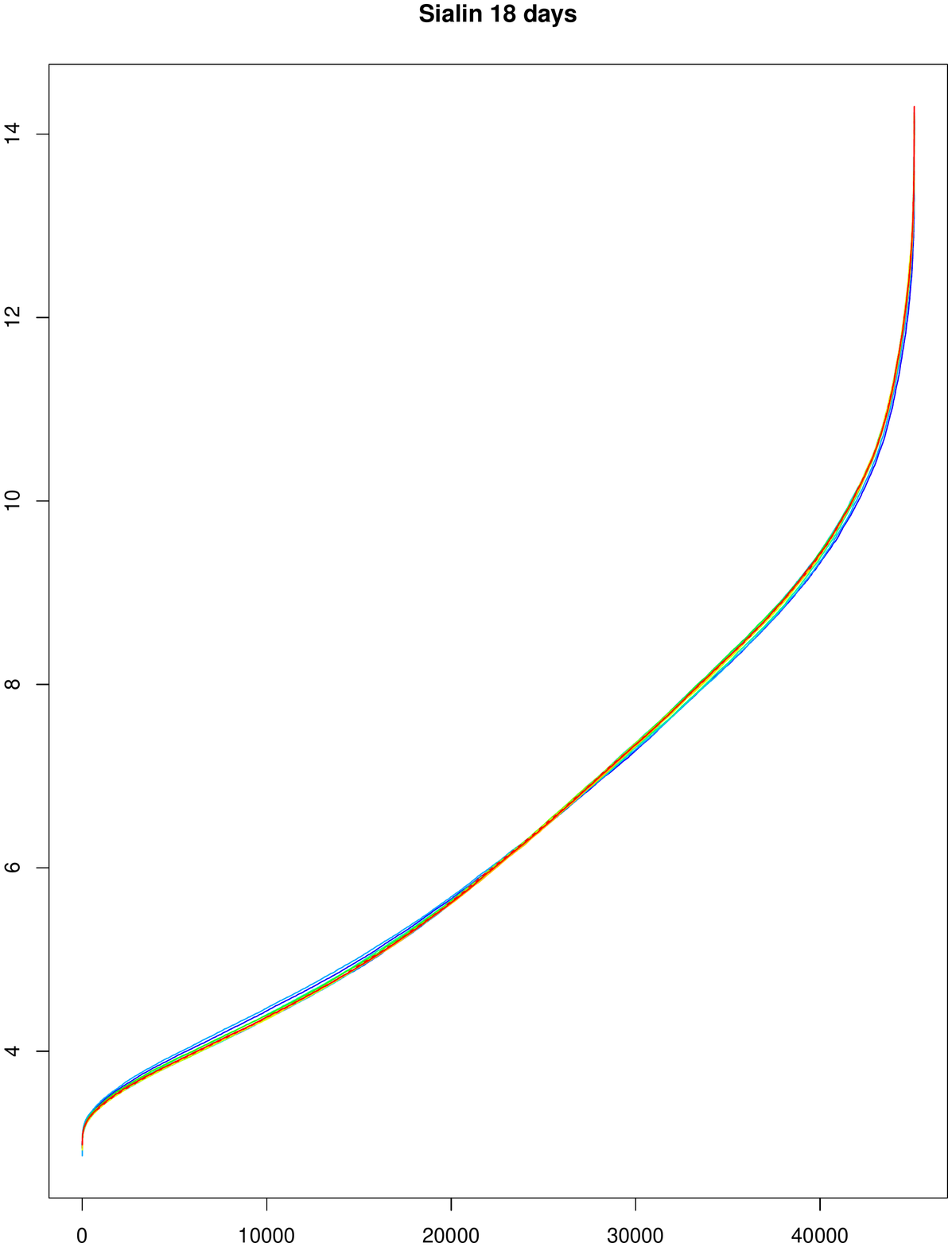}
 \includegraphics[width=.3\linewidth]{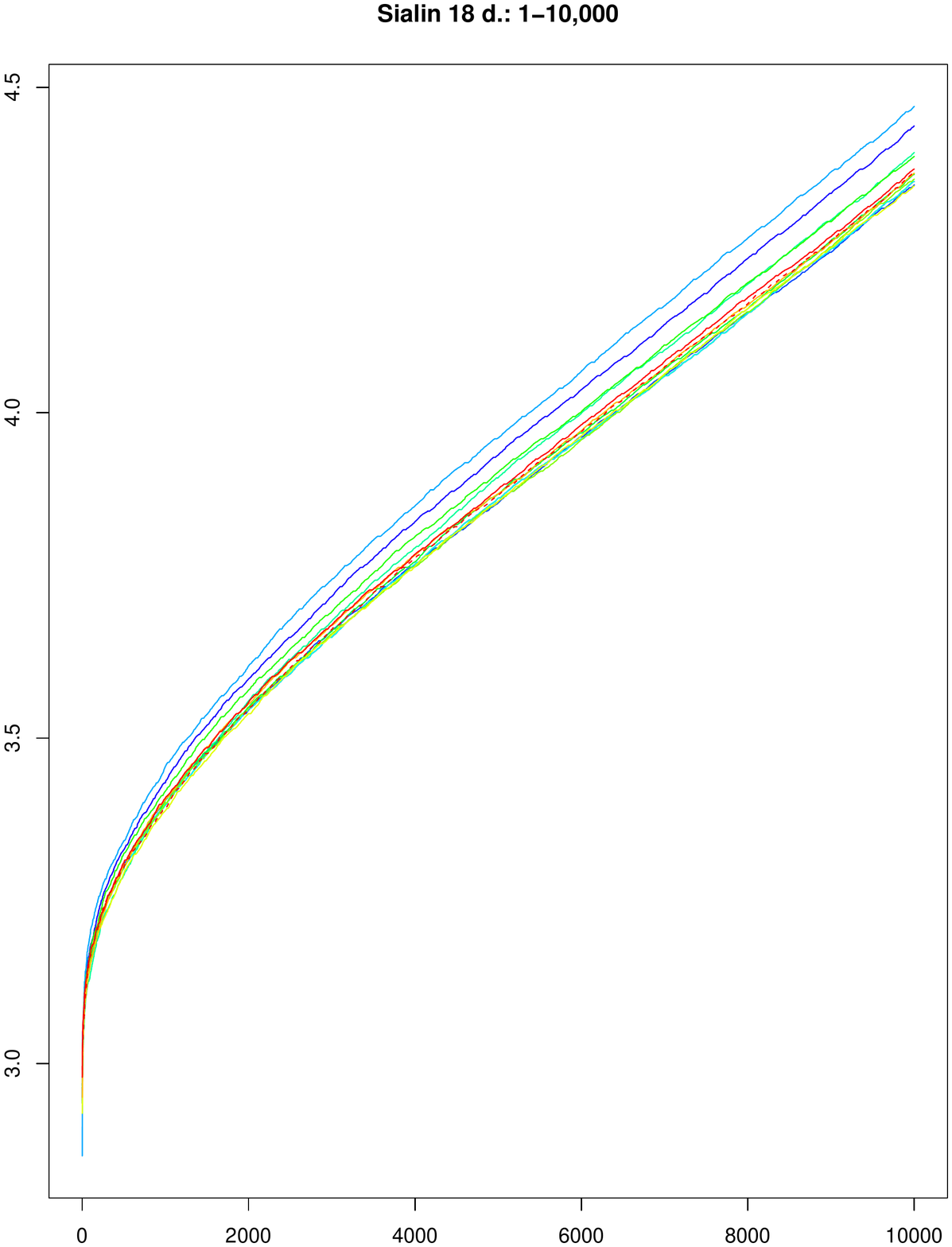}
   \includegraphics[width=.3\linewidth]{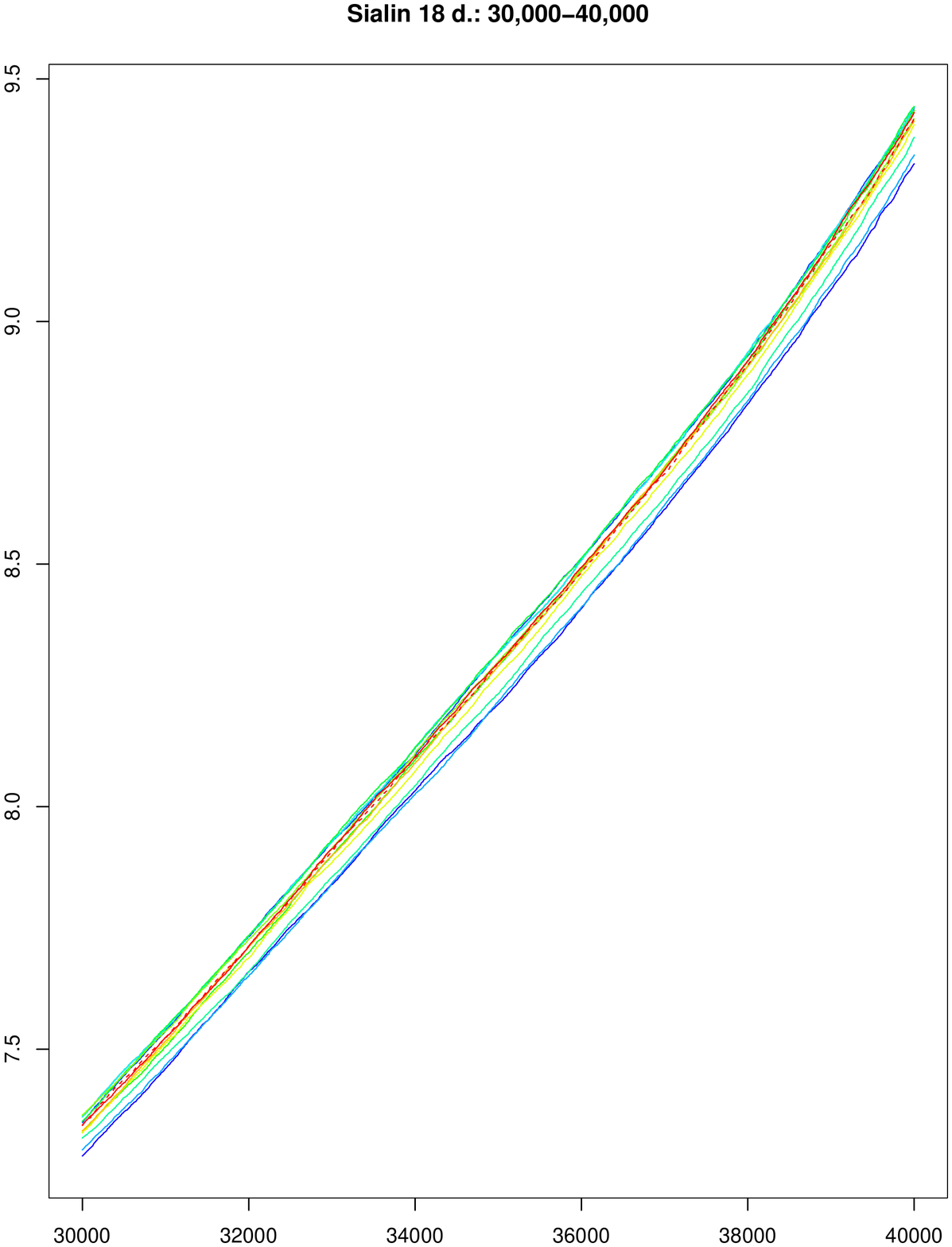}
\caption{Representation of the two $Sialin$ datasets: 12 microarrays after 6 hours (top row) of gestation and 12 after 18 days (bottom row). The data are displayed in the normalization process where the microarrays are sorted in ascending order.  The x-axis varies from 1 to 45,101 in the left column plots, where  45,101 is the total number of genes. The  central column plots of the figure represent a zoom of the first part of the domain, from 1 to 10,000.  The right column plots display  a zoom of a central part of the domain that goes from 30,000 to 40,000.
The y-axis represents the intensity levels in the microarrays.
The colors stand for the statistical functional depth of the microarrays, from blue for low statistical  functional depth to red for high statistical functional depth, through green and yellow.}
\label{JyJ}
\end{figure}

\begin{figure}[!htbp]
\includegraphics[width=.3\linewidth]{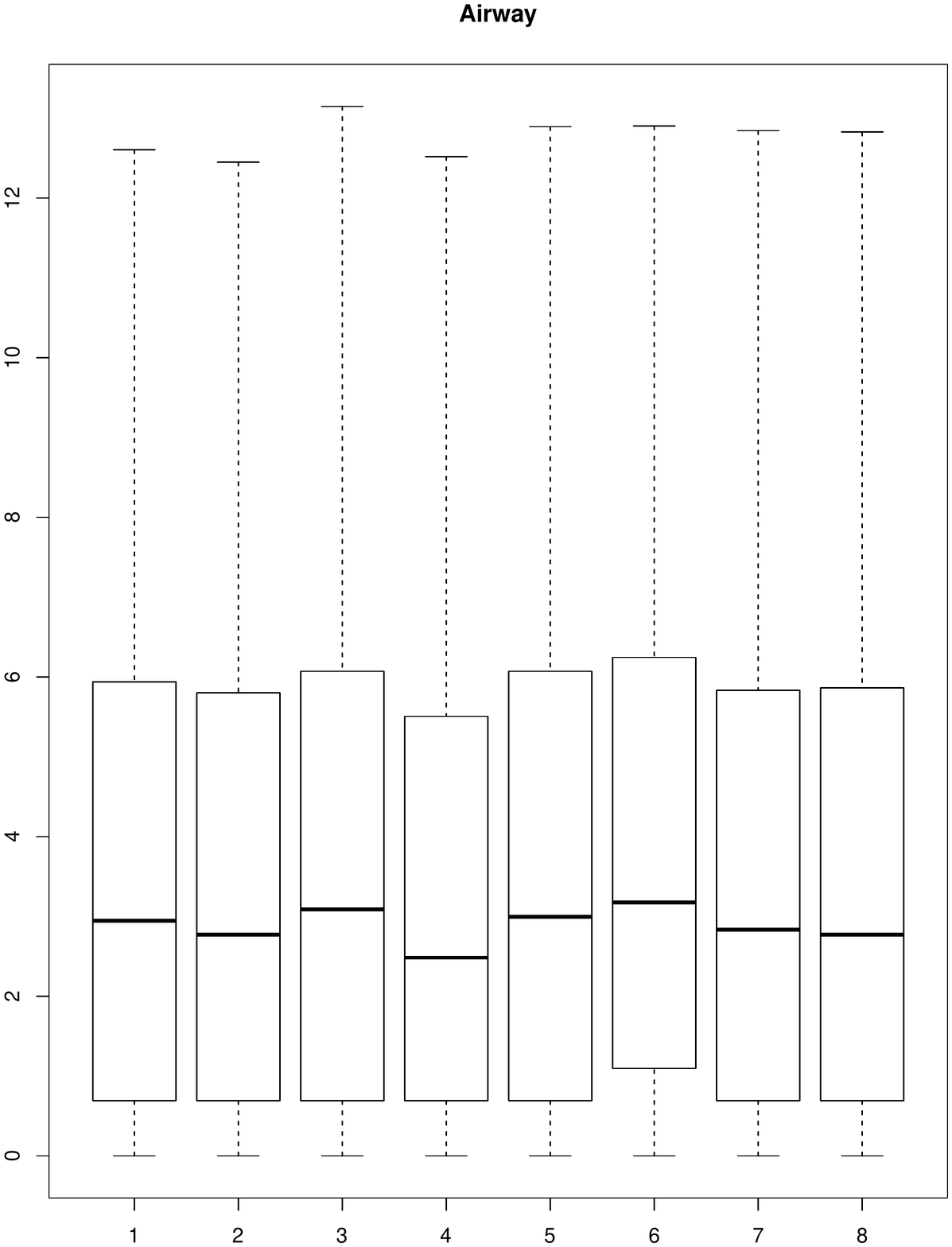} 
  \includegraphics[width=.3\linewidth]{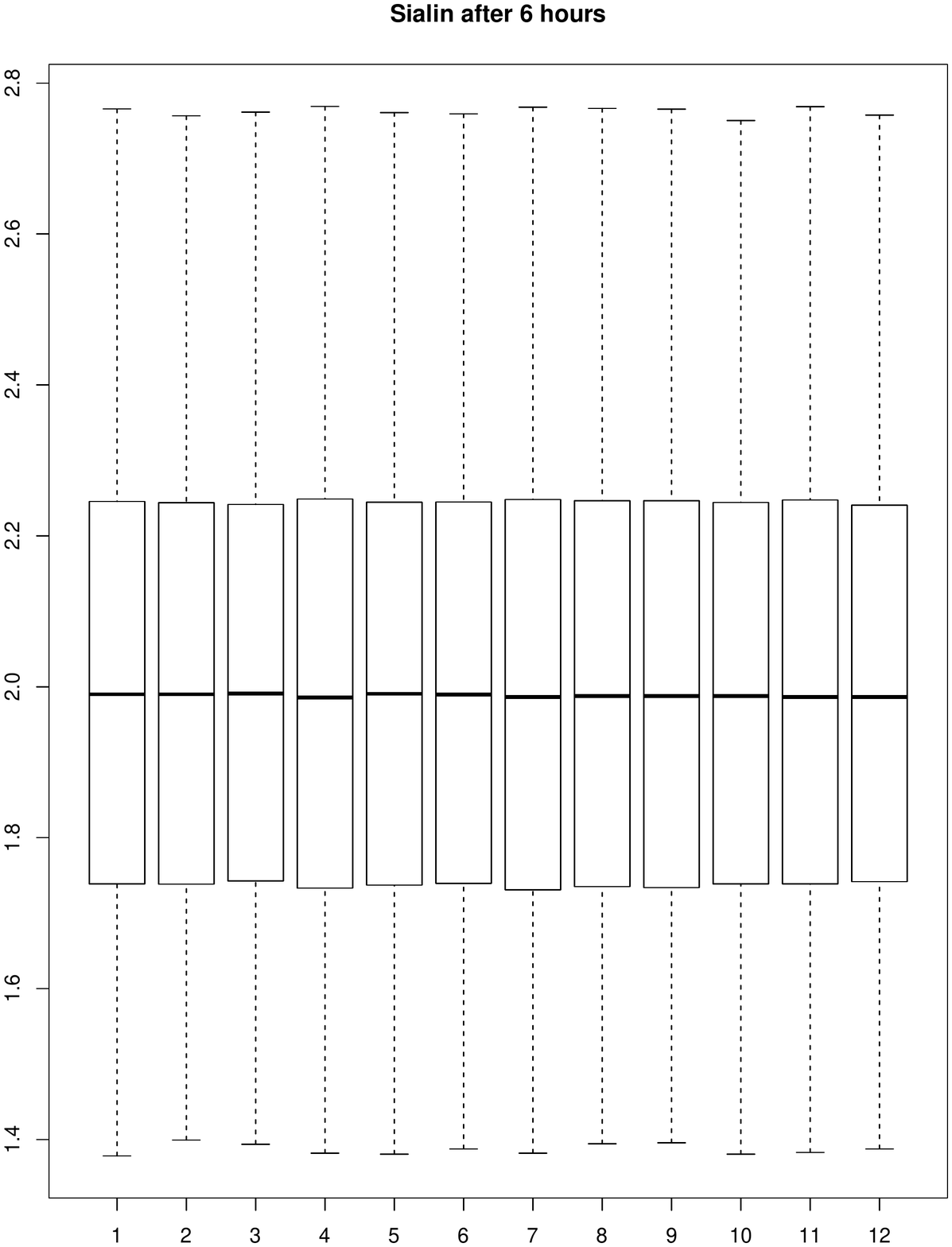}
 \includegraphics[width=.3\linewidth]{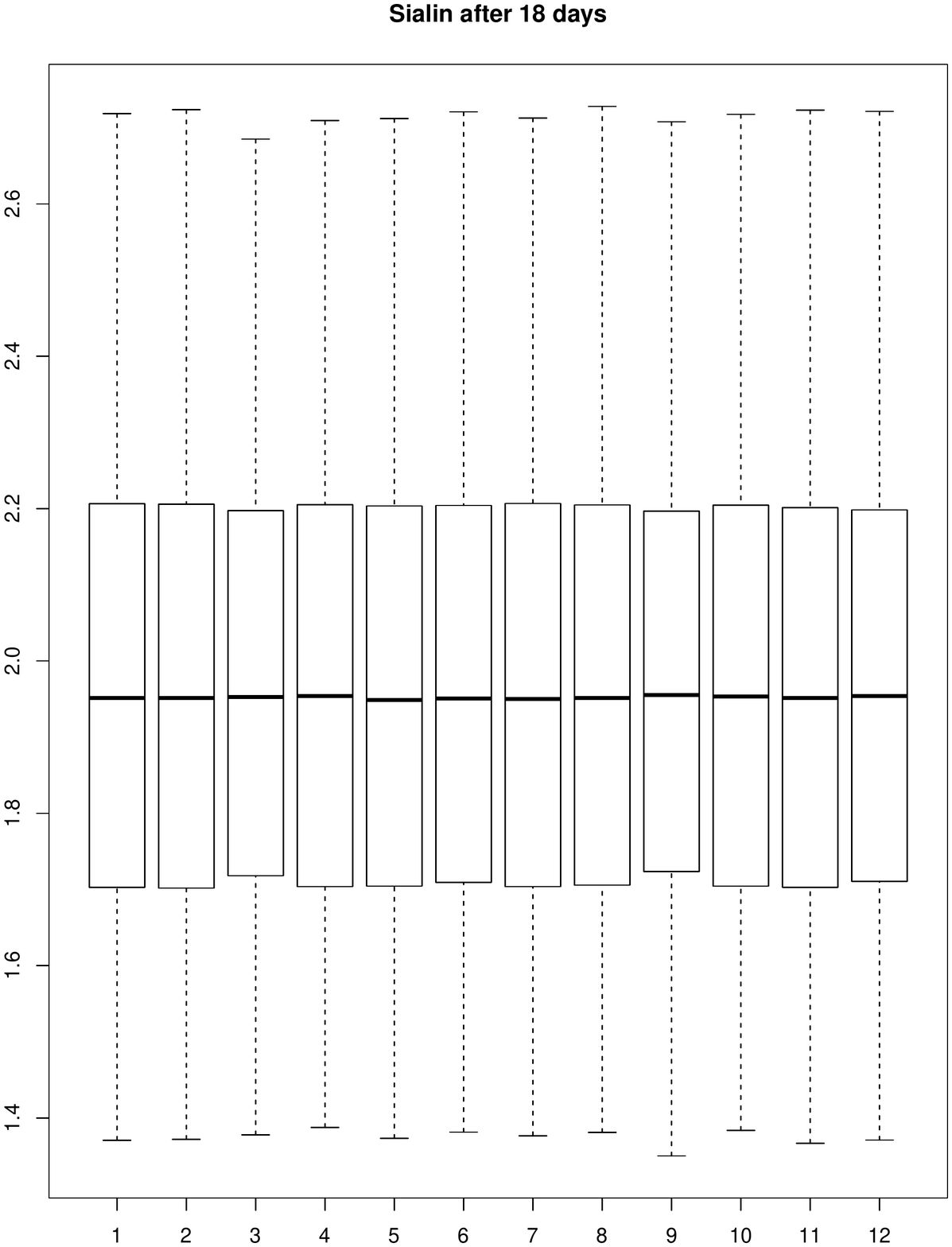}
   \includegraphics[width=.3\linewidth]{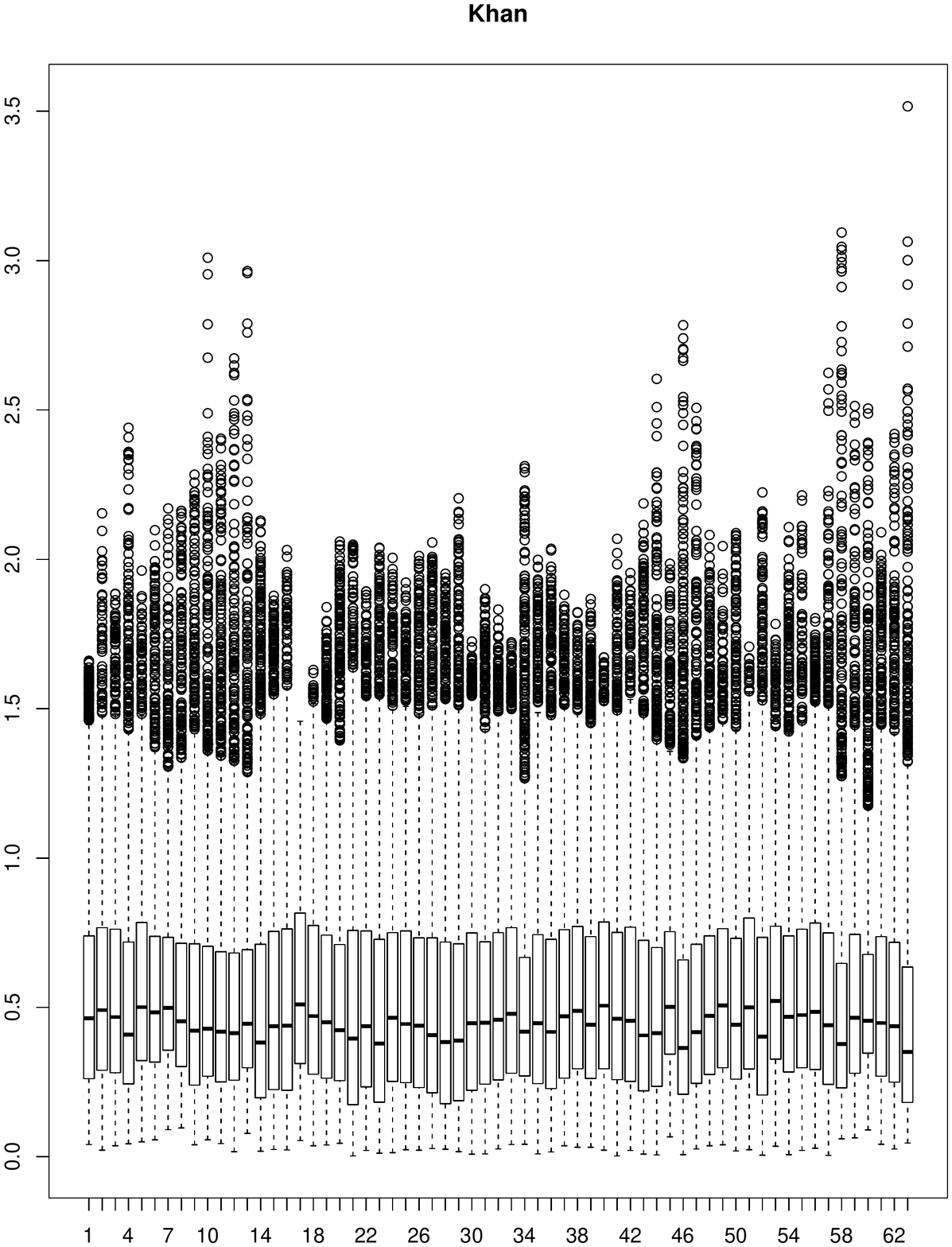}
 \includegraphics[width=.3\linewidth]{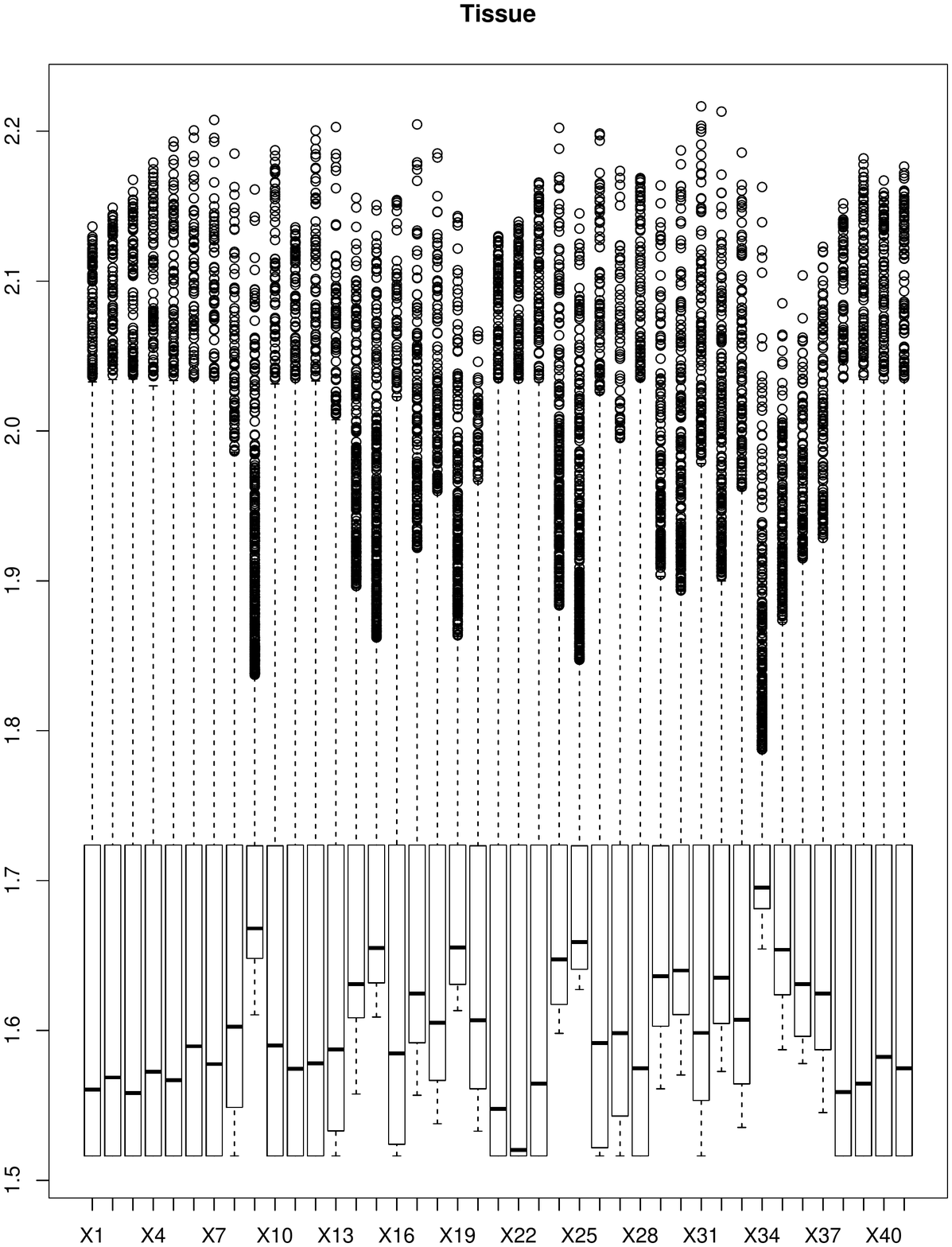}
\caption{Boxplot of the $log(\cdot+1)$ transform of the five studied datasets before the normalization step. RNA-seq Airway dataset (top left panel),  Sialin dataset after 6 hours of gestation (top central panel) and after 18 days  (top right panel), Kahn dataset (bottom left panel) and Tissue dataset (bottom right panel). The x-axis represents the RNA-seq/microarrays of each of the datasets; with the number identifiers corresponding to those in Tables \ref{names1} and  \ref{names2}.
The y-axis represents the $log(\cdot+1)$ transform of the counts, for the Airway dataset, and of the intensity levels for the Sialins, Kahn and Tissue datasets.}
\label{boxBefore}
\end{figure}

\begin{figure}[!htbp]
\includegraphics[width=.3\linewidth]{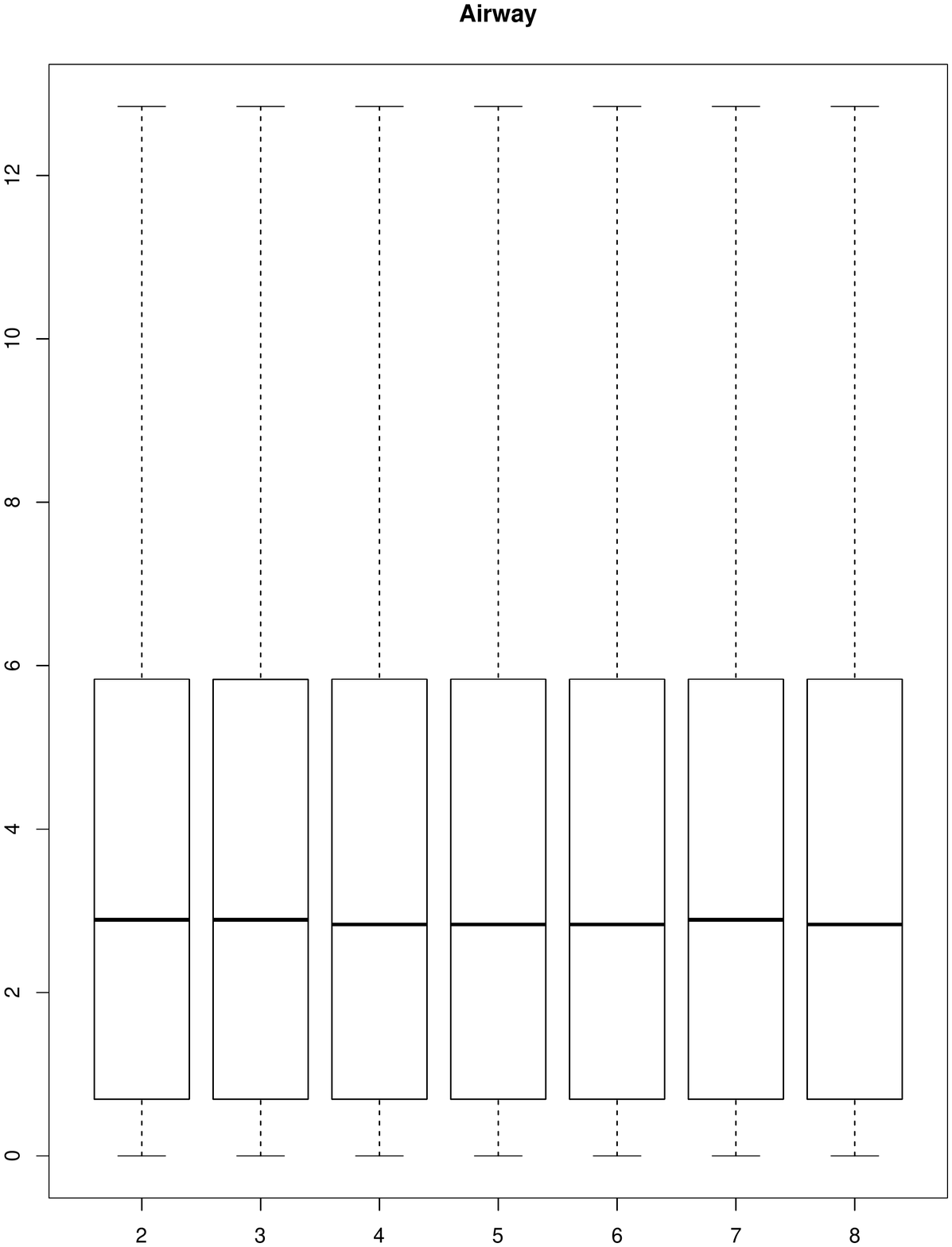} 
 \includegraphics[width=.3\linewidth]{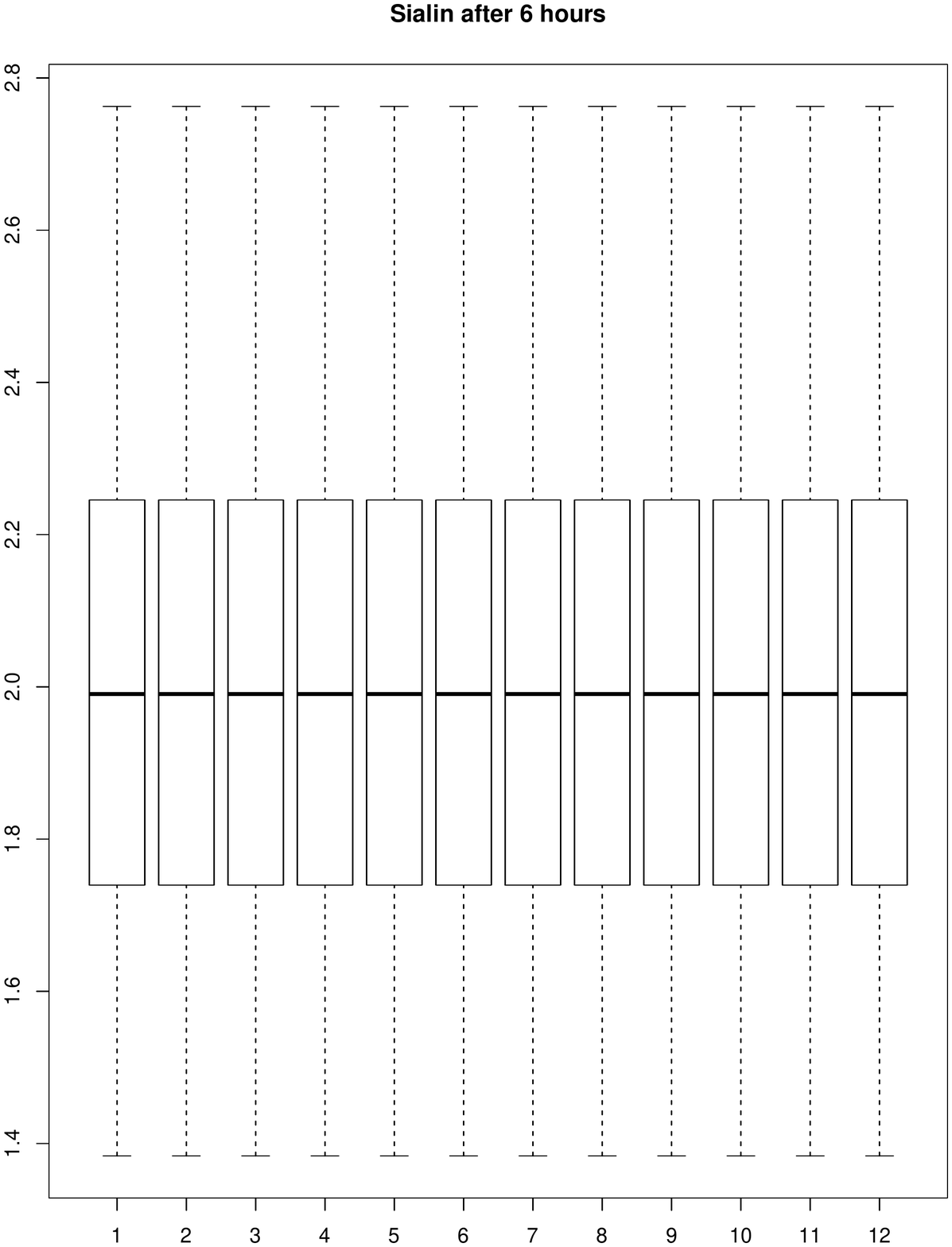}
 \includegraphics[width=.3\linewidth]{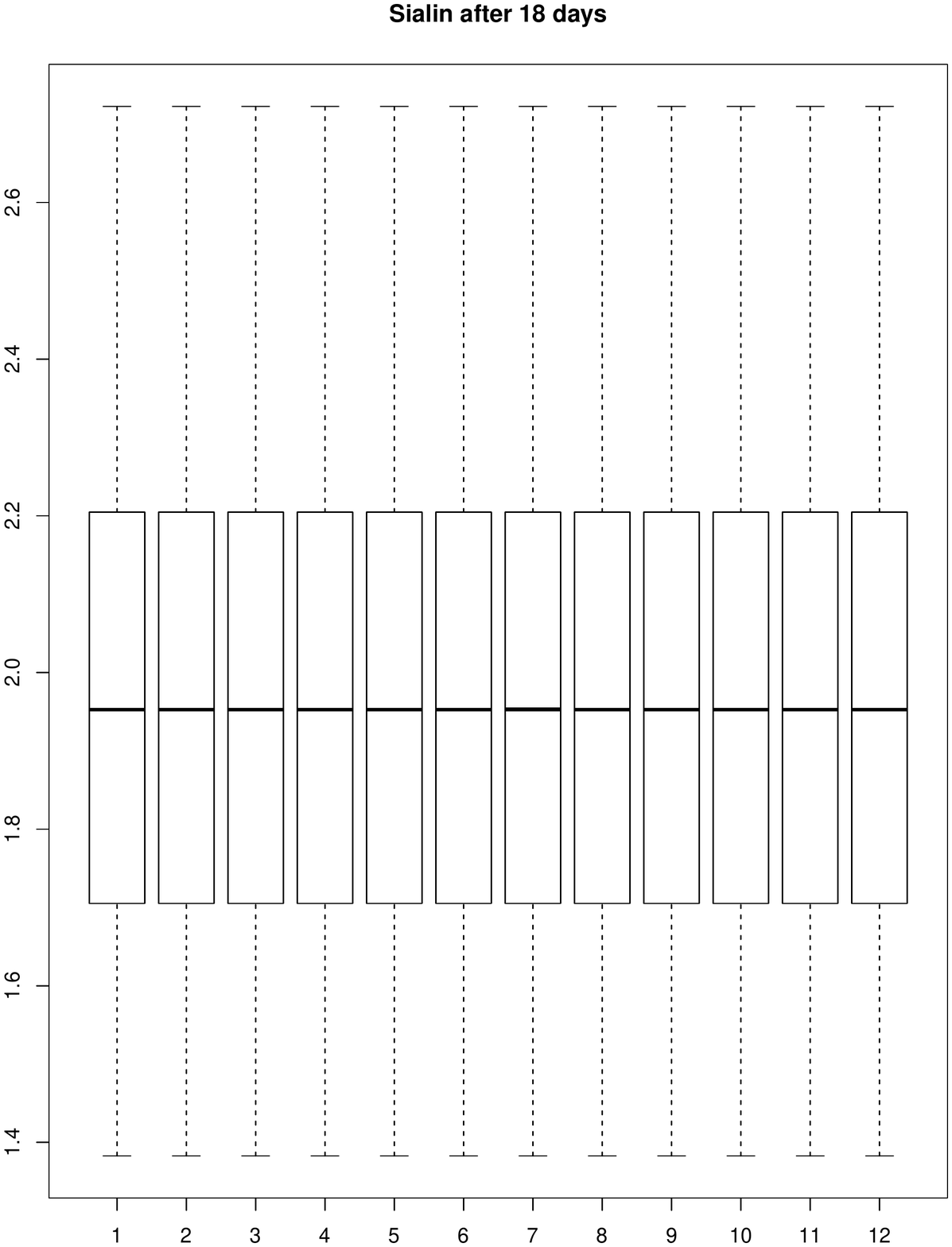}
   \includegraphics[width=.3\linewidth]{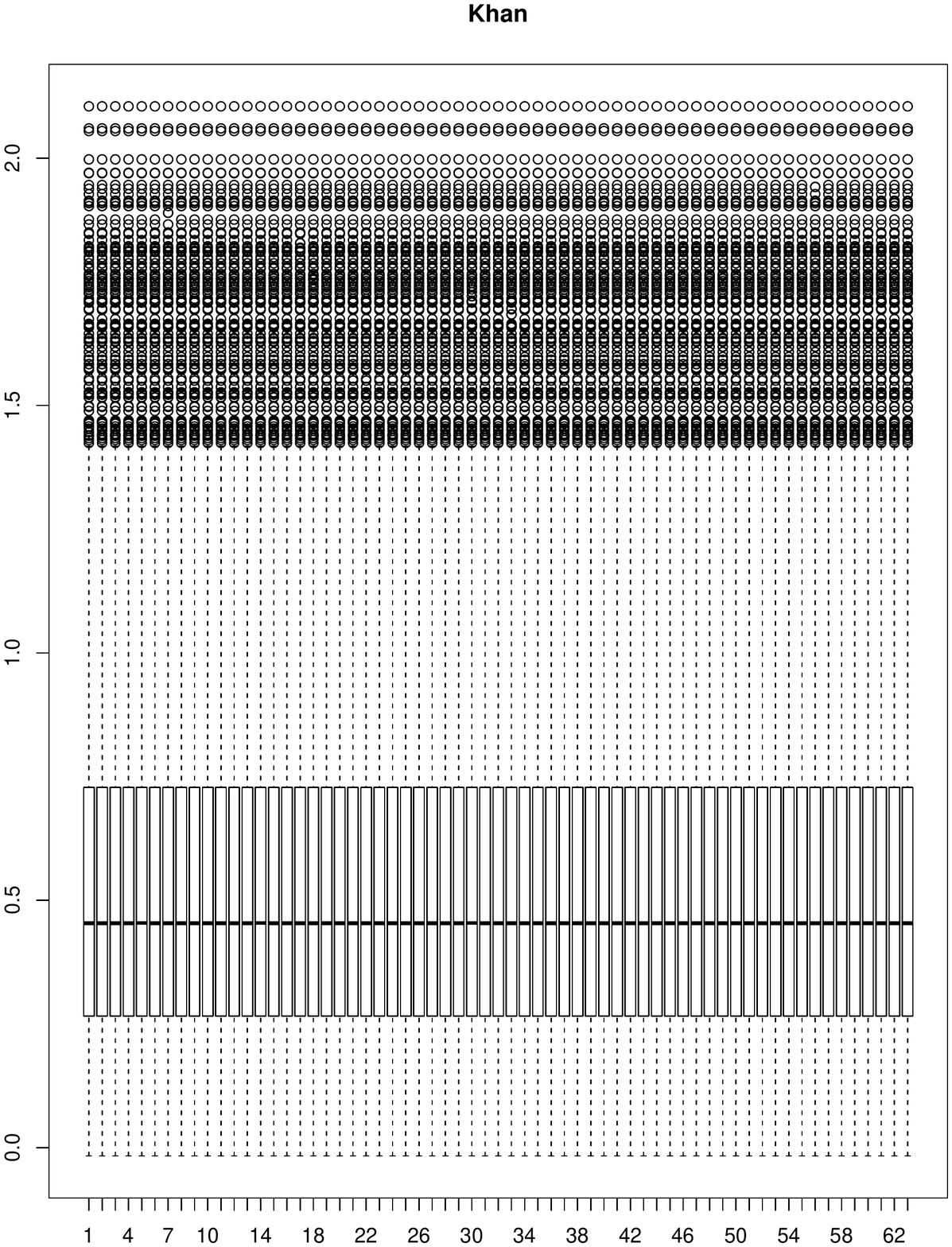}
 \includegraphics[width=.3\linewidth]{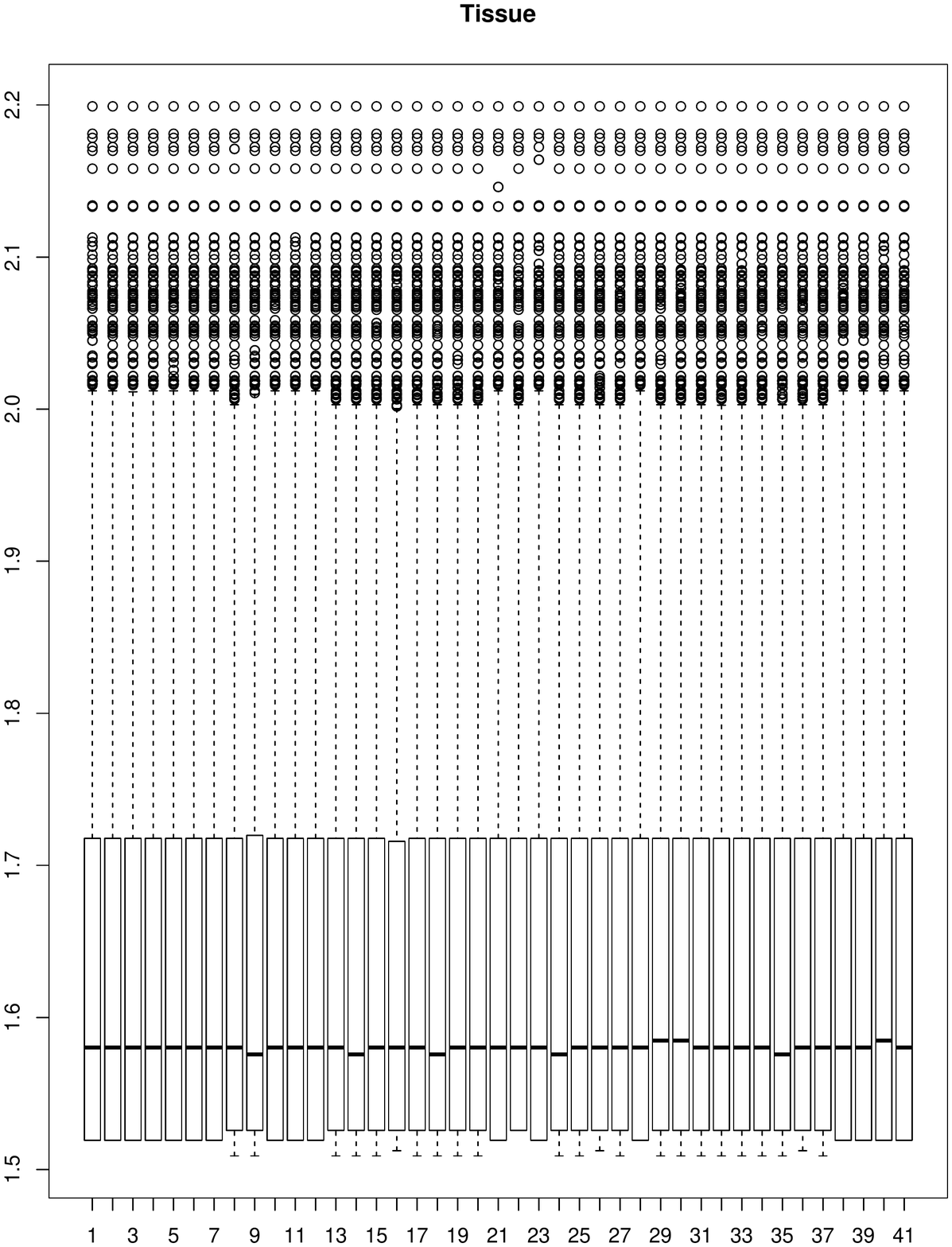}
\caption{Boxplot of the $log(\cdot+1)$ transform of five studied datasets after statistical functional depth based normalization step. RNA-seq Airway dataset (top left panel),  Sialin dataset after 6 hours of gestation (top central panel) and after 18 days  (top right panel), Kahn dataset (bottom left panel) and Tissue dataset (bottom right panel). The x-axis represents the RNA-seq/microarrays of each of the datasets; with the number identifiers corresponding to those in Tables \ref{names1} and  \ref{names2}.
The y-axis represents the $log(\cdot+1)$ transform of the counts, for the Airway dataset, and of the intensity levels for the Sialins, Kahn and Tissue datasets.}
\label{box}
\end{figure}

Through this section we work with a variety of  datasets, which we introduced in Section \ref{intro}: the airway dataset, two sets we named $Sialin$ and two known datasets, Khan and Tissue.
Due to the reader might be unfamiliar with he $Sialin$ datasets, 
we show a representation of them in the step of the normalization process where the columns have being sorted. 
Thus, in the left column plots of Figure \ref{JyJ}  we have displayed on the top row the sorted data after 6 hours of gestation  and  on the bottom row after 18 days. The  central column plots of the figure represent a zoom of the first part of the domain and the right column plots a zoom of a central part of the domain. Making use of the colors of the rainbow, we represent the statistical functional depth of the curves, from blue for low statistical  functional depth to red for high statistical functional depth.
Once the normalization based on statistical functional depth has being carried out, to show the adequacy  of the procedure,  we have done a boxplot   of the $log (\cdot+1)$ transform  of the resulting datasets, which can be seen in  Figure \ref{box}. To appreciate the effect of the normalization step, we have included in Figure \ref{boxBefore} the boxplot of the $log (\cdot+1)$ transform of the datasets previous to the normalization step.

In what follows, we compare empirically the proposed analytical procedure, described in Section \ref{om}, with the two mentioned  graphical procedures.

\subsection{Airway dataset}
We apply the analytical proposed procedure on  the $Airway$ dataset and obtain that it contains an outlier. This can be observed in Table \ref{TA} where each pair of RNA-seq's is displayed following their statistical depth order. Thus, the pair $(5,2)$ is the statistical least deep one and the pair $(8, 7)$ the statistical deepest one. Below each pair of RNA-seq's, it is shown the $L_2$ (euclidean) distance between the elements in the pair. A pair contains a potential outlier when this distance is larger than the   benchmark for outliers, which is also displayed in the table.  For instance,  the distance between the elements of the statistical least deep pair of RNA-seq's,  $484,974.7,$ is just above the  benchmark for outliers, which is $484,923.1.$  Between the two elements of the statistical least deep pair, the RNA-seq with identification number  5 is the one consider as potential outlier because it has the larger $L_2$ (euclidean) distance to the average of statistical deepest elements, with identification numbers 8 and 7.

\begin{table}[!htb]
\begin{center}
\begin{tabular}{r|cccc}
& \multicolumn{4}{c}{Airway} 
\\
\hline
&&&&
\\
pairs of gene &     {\bf 5}&     1&      2&     8\\
expressions &   4&      6&     3&      7\\
&&&&
\\
distance intra-pair& {\bf484,974.7 }&471,420.6 &346,877.9& 242,326\\
&&&&
\\
outlier's benchmark & \multicolumn{4}{c}{484,923.1} \\
Tukey's constant& \multicolumn{4}{c}{1.2}
\\
\hline
\end{tabular}
\end{center}
\caption{The first two rows are the pairs of RNA-seq's of the Airway dataset ordered from statistical least deep (left) to deepest one (right). The third row contains the distance between the two elements of each corresponding pair, i.e., the distance between  RNA-seq's with identification number  5 and 4 is $484,974.7.$  The fourth row is the value $G\cdot d(Q_3 ,Q_1)$ of Inequality \ref{mild3}, the benchmark to detect a pair of gene expressions as outliers. The fifth row is the value of $G.$ The first intra-pair distance and the RNA-seq identifier 5 are in bold to represent it is a potential outlier.}\label{TA}
\end{table}

\begin{figure}[!htbp]
\includegraphics[width=.3\linewidth]{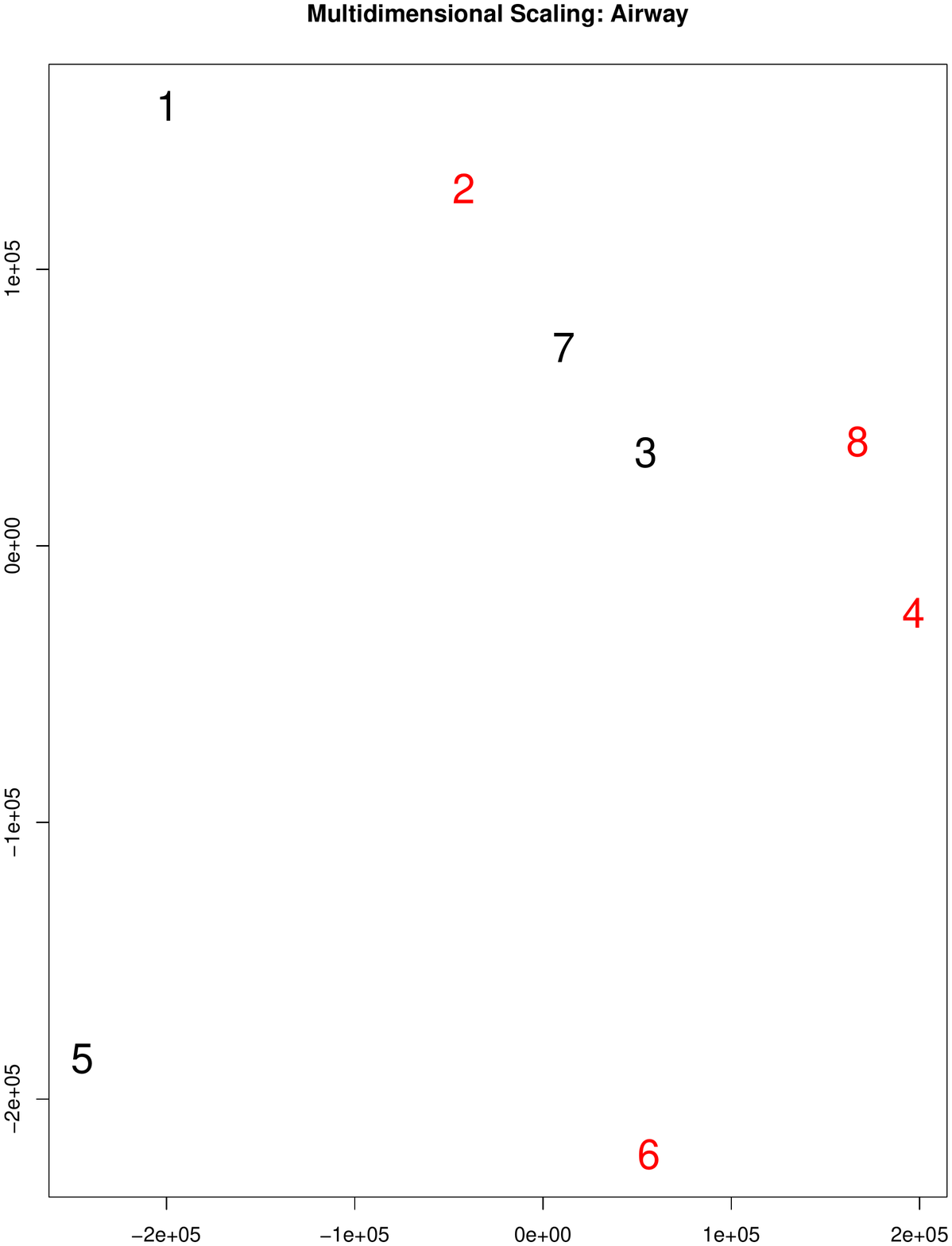}
 \includegraphics[width=.3\linewidth]{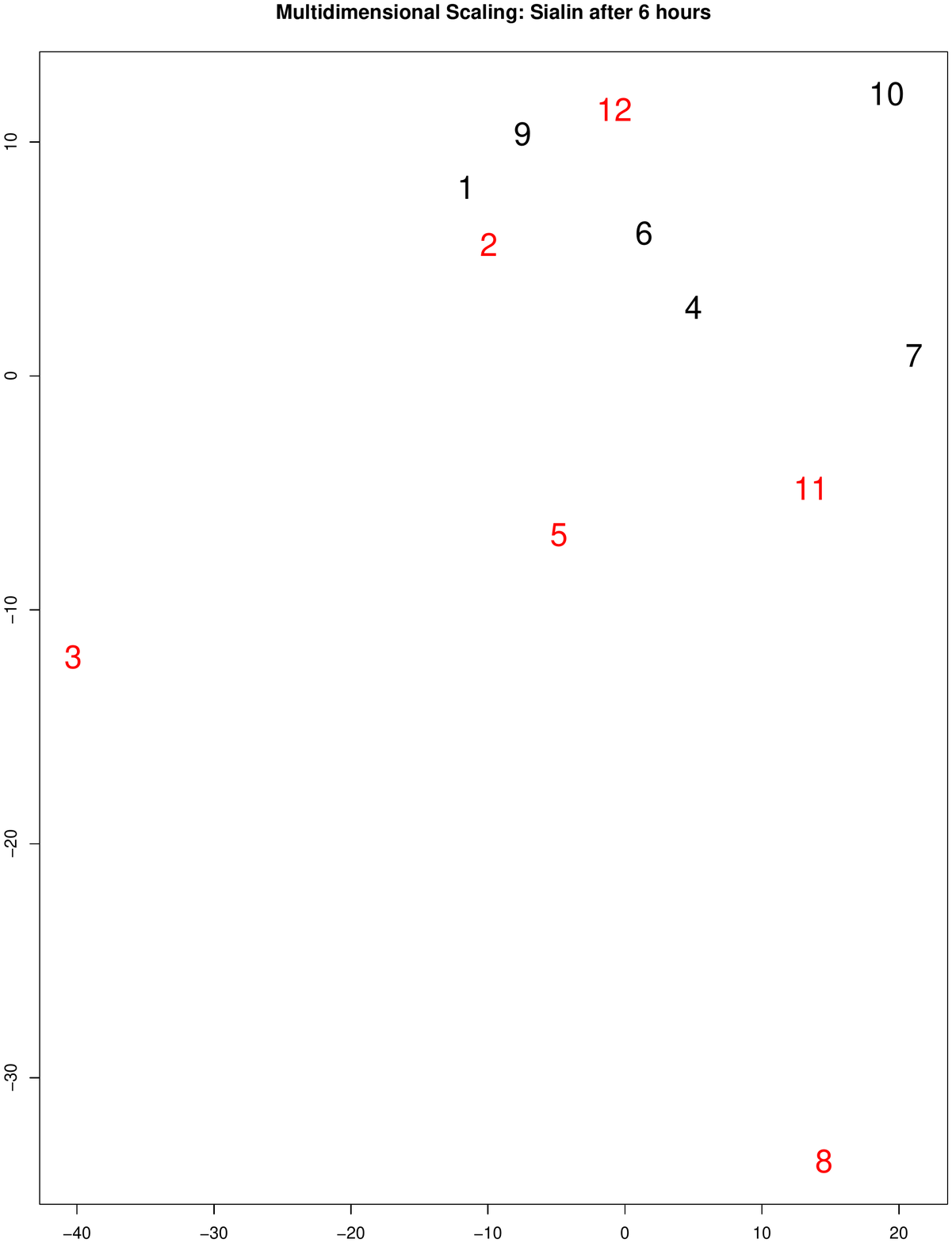}
  \includegraphics[width=.3\linewidth]{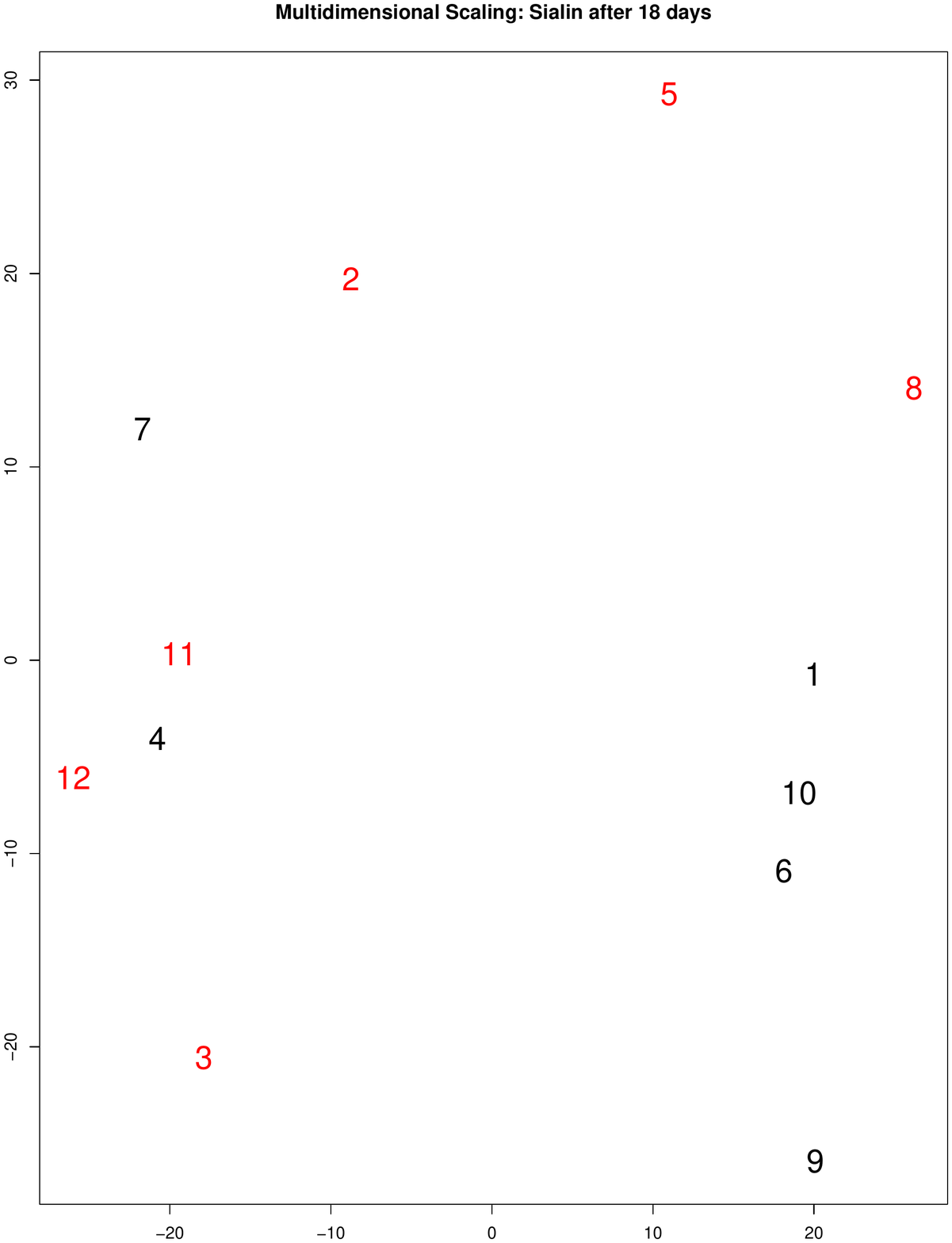}
   \includegraphics[width=.3\linewidth]{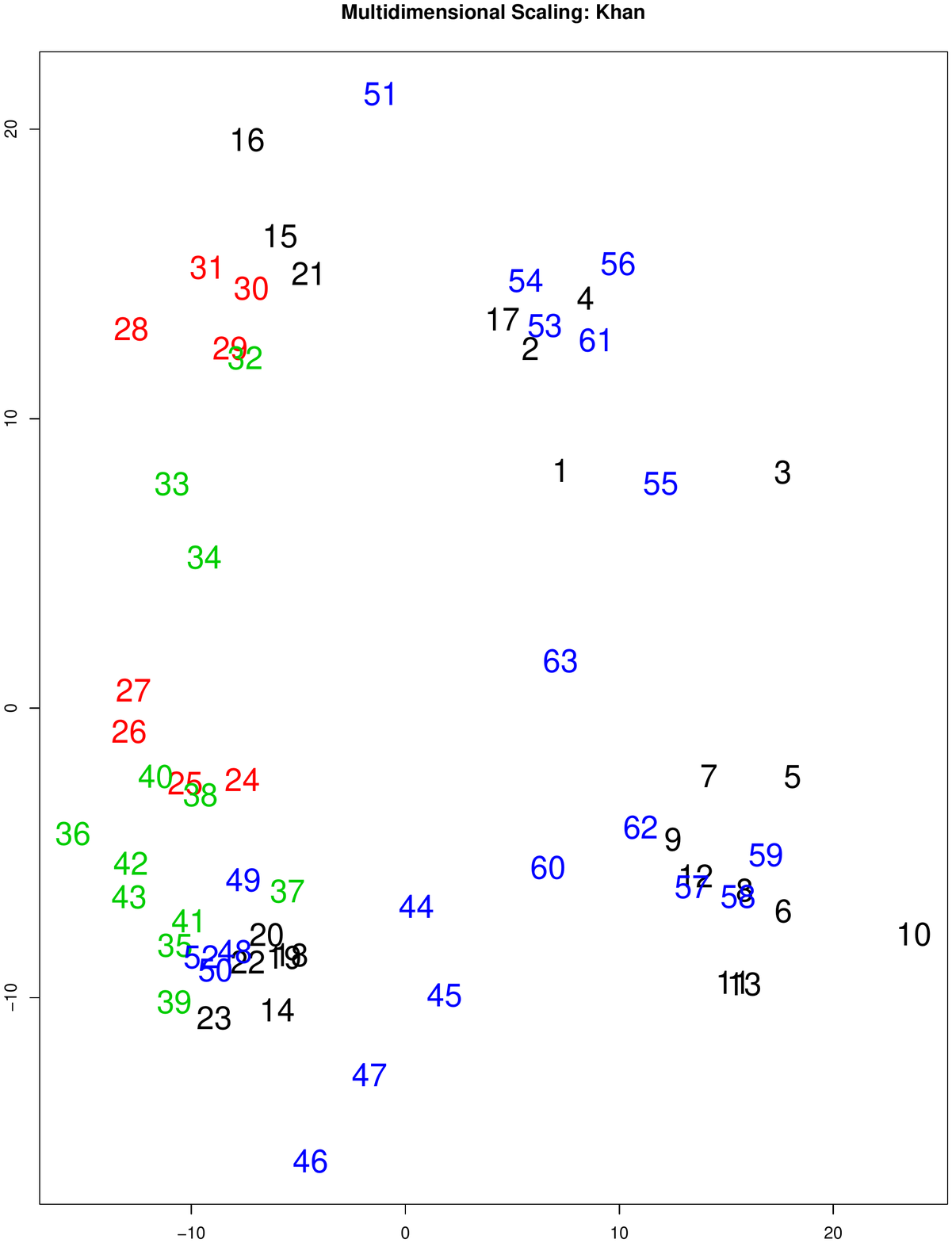}
 \includegraphics[width=.3\linewidth]{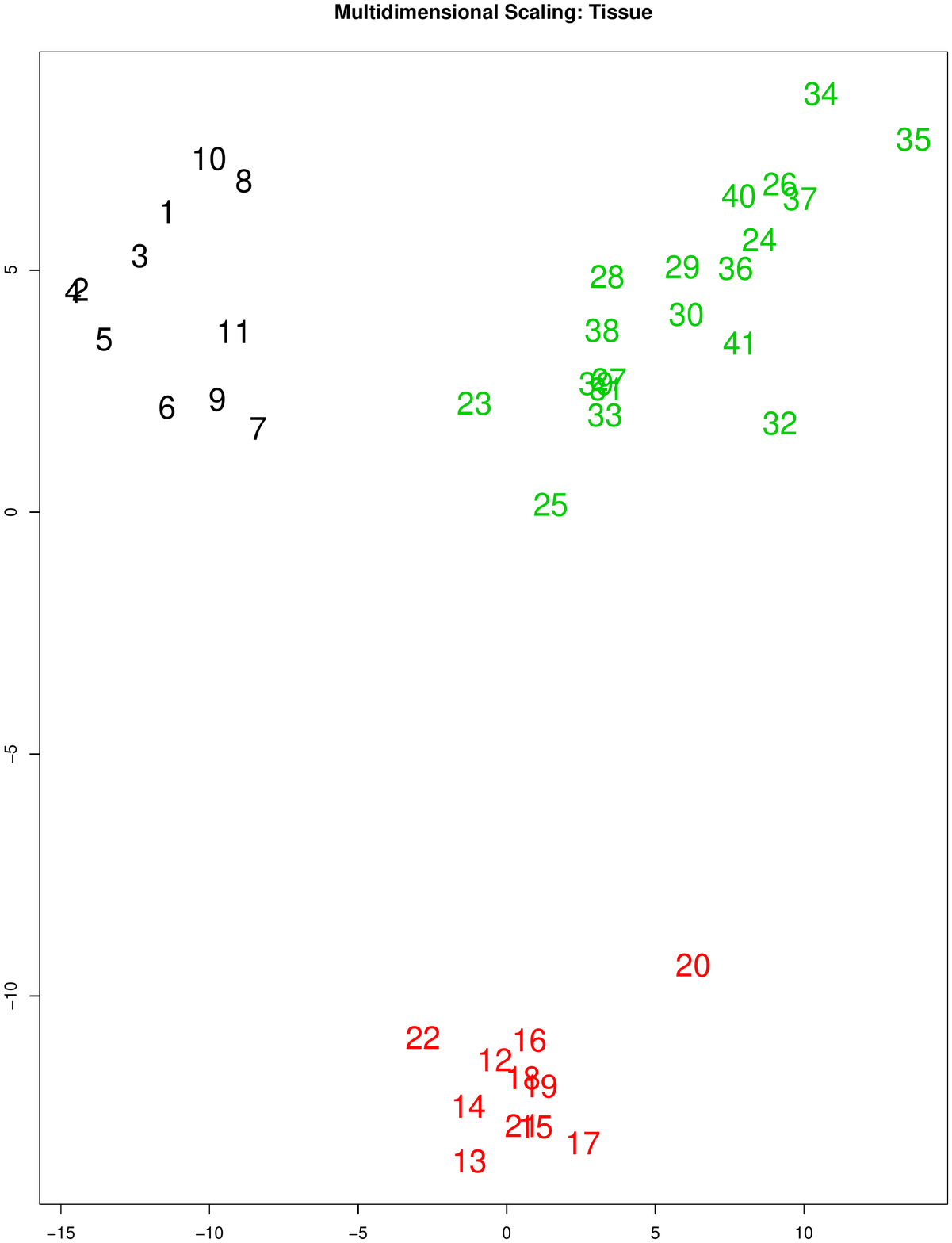}
\caption{Multidimensional scaling plots of the datasets: Airway dataset (top left panel), Sialin datasets (top central and left panel), Khan dataset (bottom left panel) and Tissue dataset (bottom  right panel).  For the Sialin datasets, the data from the study after 6 hours of gestation is displayed  in the top central panel and the data from the study after 18 days in the top left panel. 
The multidimensional scaling results are a two dimensional representation of each microarray that shows the differences among them.
The x-axis values correspond to the first coordinate of that representation and the y-axis values to the second.
The numbers in each plot are the identifiers of each RNA-seq or microarray as displayed in Tables \ref{names1} and \ref{names2}.  
According to this methodology, those far away from the cluster are potential outliers, for instance, in the Airway dataset those RNA-seq identified with number 5 and 6.
The colors represent the class to which the identification value belongs: black (class1), red (class 2), green (class 3) and blue (class 4).}
\label{A}
\end{figure}

The  two graphical procedures are: one in the top left plot of Figure \ref{A}, multidimensional scaling; the other,  in the top left plot Figure 2
from  the Supplementary material, spectral map analysis.
They 
show that the RNA-seq's 5 and 6 are distant from the others; but it is not clear from these graphical procedures where they are distant enough to be consider as outliers. In fact, these two RNA-seq's belong to the two statistical least deep pair of RNA-seq's and each of them belongs to a different  class. According to the analytical procedure, RNA-seq 6 is an outlier when the computation of our interquartile distance is restricted to the class it belongs.  This is observable from Table \ref{Ac}, from where it can also be seen that RNA-seq 5 is not a potential outlier when restricted to its class.

\begin{table}[!htb]
\begin{center}
\begin{tabular}{r|cc|cc}& \multicolumn{4}{c}{Airway}\\
& \multicolumn{2}{c}{class 1}& \multicolumn{2}{c}{class 2}
\\
\hline
&&&&
\\
pairs of gene &     5    &  1 &{\bf 6}&4\\
expressions &    3     & 7&2&8\\
&&&&
\\
distance intra-pair& 409,994 & 307,657.2&{\bf401,539.5 }&171,404.9\\
&&&&\\
outlier's benchmark & \multicolumn{2}{c|}{425,344}& \multicolumn{2}{c}{339,354.4} \\
Tukey's constant& \multicolumn{2}{c|}{1.2}& \multicolumn{2}{c}{1.2}\\
\hline
\end{tabular}
\end{center}
\caption{Inside each class, the first two rows are the two pairs of RNA-seq's of the Airway dataset ordered by statistical depth: least deep (left) and deepest (right). The third row contains the distance between the two elements of each corresponding pair.  The fourth row is the value $G\cdot d(Q_3 ,Q_1)$ of Inequality \ref{mild3}, the benchmark to detect a pair of gene expressions as outliers. The fifth row is the value of $G$ from Table \ref{TA}. The first intra-pair distance of class 2 and the RNA-seq identifier 6 are in bold to represent it is a potential outlier with respect to its class.}\label{Ac}
\end{table}

\subsection{Sialin datasets}

Using the analytical proposed procedure on each of the $Sialin$ datasets, we obtain that the microarray with identification number 3 is consistently depicted as an outlier.
 This can be observed in Table \ref{TJ} where the four less deepest pairs of microarrays  of the study after 6 hours of gestation and  of the study after 18 days are displayed. Additionally, when we restrict ourselves to each class to compute the distance corresponding to the interquartile range, microarray with identification number 3 is also depicted as an outlier in the study after 6 hours of gestation, see Table \ref{TJc}.
  Regarding the  two graphical procedures, they point to  microarray number 3 being a potential outlier  in the study after 6 hours of gestation, taking the whole set of data or just the class to which it belongs. However, the graphical procedures do not indicate clearly that it is an outlier in the study after 18 days.
  Furthermore, the graphical procedures may also indicate that microarray number 8 is a potential outlier in the study after 6 hours of gestation while Table  \ref{TJ} shows that it belongs two the second statistical less deep pair of microarrays with a distance between its elements of $58.1$ that is smaller than the  benchmark for outliers of $67.6.$
The multidimensional scaling procedure is displayed in the top row of Figure \ref{A}; the central plot is for the study after 6 hours and the right plot for the study after 18 days. Analogously, the spectral map analysis is  displayed in the top central and right panels of  Figure 2 
from  the Supplementary material; the left plot is for the study after 6 hours and the right plot for the study after 18 days.

\begin{table}[!htb]
\begin{center}
\begin{tabular}{r|cccc|cccc}
& \multicolumn{8}{c}{Sialin}
\\
& \multicolumn{4}{c}{after 6 hours}& \multicolumn{4}{c}{after 18 days}
\\
\hline
&&&&&&&&
\\
pairs of gene & {\bf3}&  9&  4&  7&{\bf3}& 11& 12&  7
\\
expressions &  1&  8& 12&  1&8&  5&  9& 10
\\
&&&&&&&&
\\
distance intra-pair& {\bf68.1} &58.1 &54.5 &51.1&{\bf69.3 }&66.5 &66.0 &61.1
\\
&&&&&&&&\\
outlier's benchmark & \multicolumn{4}{c|}{67.6}& \multicolumn{4}{c}{69.1} \\
Tukey's constant& \multicolumn{4}{c|}{1.3}& \multicolumn{4}{c}{1.1}\\
\hline
\end{tabular}
\end{center}
\caption{The first two rows are the pairs of gene expressions identifiers of the Sialin datasets, where each dataset is  ordered from statistical least deep (left) to deepest one (right). The third row contains the distance between the two elements of each corresponding pair.  The fourth row is the value $G\cdot d(Q_3 ,Q_1)$ of Inequality \ref{mild3}, the benchmark for each dataset to detect a pair of gene expressions as outliers. The fifth row is the value of $G$ obtained for each dataset. The first intra-pair distance of each dataset and the gene expression  identifiers 3 are in bold to represent they are  potential outliers.}\label{TJ}
\end{table}

\begin{table}[!htb]
\begin{center}
\begin{tabular}{r|cc|cc||cc|cc}
& \multicolumn{8}{c}{Sialin}
\\
& \multicolumn{4}{c}{after 6 hours}& \multicolumn{4}{c}{after 18 days}
\\
& \multicolumn{2}{c}{class 1}& \multicolumn{2}{c||}{class 2}& \multicolumn{2}{c}{class 1}& \multicolumn{2}{c}{class 2}
\\
\hline
&&&&&&&&
\\
pairs of gene & 4  &7   &{\bf 3}& 12 & 7&4  &3&11
\\
expressions & 9 &  1& 8& 5&9 &10  &8&5
 \\
&&&&&&&&
\\
distance intra-pair& 53.3 &51.1&{\bf 66.1} &50.1&65.3&60.9&69.3 &66.5
\\
&&&&&&&&\\
outlier's benchmark & \multicolumn{2}{c|}{65.4}&\multicolumn{2}{c||}{64.1}&
 \multicolumn{2}{c|}{66.1}&\multicolumn{2}{c}{72.3} \\
Tukey's constant& \multicolumn{2}{c|}{1.3}&\multicolumn{2}{c||}{1.3}&
  \multicolumn{2}{c|}{1.1}& \multicolumn{2}{c}{1.1}
\\
\hline
\end{tabular}
\end{center}
\caption{Inside each class and dataset, the first two rows are the two pairs of gene expressions of the Sialin datasets ordered by statistical depth: least deep (left) and deepest (right). The third row contains the distance between the two elements of each corresponding pair.  The fourth row is the value $G\cdot d(Q_3 ,Q_1)$ of Inequality \ref{mild3}, the benchmark to detect a pair of gene expressions as outliers. The fifth row is the value of $G$ from Table \ref{TJ}. The first intra-pair distance of class 2 of the Sialin dataset after 6 hours of gestation and microarray number 3 are  in bold to represent it is a potential outlier with respect to its class.}\label{TJc}
\end{table}

\subsection{Khan dataset}
The analytical procedure on the Khan data produces no outlier when the dataset is study as a whole. This can be observed in Table \ref{TK} where the eight statistical least deep pairs of microarrays are displayed. In addition, the table contains the benchmarks for outliers and the constant estimated to compute the benchmark. This can also be observed in the graphical procedures, see 
the bottom left plot of Figure \ref{A} for the multidimensional scaling procedure and  the bottom left plot of  Figure 2 
from  the Supplementary material for the spectral map analysis. 
The 63 microarrays we work with are a training set formed by four classes. Applying the analytical procedure to the first class we obtain two outliers,  microarrays number 16 and 4. If we apply the procedure to the second class, we obtain no outlier. However, applying the procedure to the third and the fourth class separately, we obtain one  outlier in each, microarray number 32 and 51 respectively. This information is displayed in Table \ref{TKc} where the  two statistical least deep pairs of each of the four classes are displayed emphasizing in bold the outliers. Notice that the outliers are from the least deep pairs of microarrays in their classes.
These findings can be seen as coherent with the graphical procedures. For instance, in the multidimensional scaling plot, microarrays 16 and 4 are in the up part of their class and microarrays 32 and 51 are plotted on the top of their respective  class.

\begin{table}[!htb]
\begin{center}
\begin{tabular}{r|cccccccc}
& \multicolumn{8}{c}{Khan}
\\
\hline
&&&&&&&&
\\
pairs of gene & 10& 58& 44& 60& 63& 57&  6& 62
\\
expressions & 16&51&  4&17&  5& 14& 33& 28
\\
&&&&&&&&
\\
distance intra-pair& 55.3 &54.3& 53.9 &52.1 &50.7 &50.2& 49.6& 48.4
\\
&&&&&&&&\\
outlier's benchmark & \multicolumn{8}{c}{55.4}\\
Tukey's constant&  \multicolumn{8}{c}{1.2}\\
\hline
\end{tabular}
\end{center}
\caption{The first two rows are the eight statistical less deep pairs of gene expressions of the Khan dataset, with the statistical least deep on the left. The third row contains the distance between the two elements of each corresponding pair.  The fourth row is the value $G\cdot d(Q_3 ,Q_1)$ of Inequality \ref{mild3}, the benchmark of the dataset to detect whether each pair of gene expressions are potential  outliers. The fifth row is the value of $G$ obtained for this dataset.}\label{TK}
\end{table}

\begin{table}[!htb]
\begin{center}
\begin{tabular}{r|cc|cc|cc|cc}
& \multicolumn{8}{c}{Khan}
\\
& \multicolumn{2}{c}{class 1}& \multicolumn{2}{c}{class 2}& \multicolumn{2}{c}{class 3}& \multicolumn{2}{c}{class 4}
\\
\hline
&&&&&&&&
\\
pairs of gene & 10  &  22&24& 27&{\bf 32}&39
    &{\bf 51}&63
\\
expressions &{\bf 16}& {\bf 4}& 28& 29&37 &33
    &58&49
    \\
&&&&&&&&
\\
distance intra-pair& {\bf55.3} &{\bf53.2}&37.2 &36.7&{\bf 42.1} &40.8
 & {\bf 54.3 }&47.6
\\
&&&&&&&&\\
outlier's benchmark & \multicolumn{2}{c|}{52.8}&  \multicolumn{2}{c|}{40.9}&  \multicolumn{2}{c|}{41.1}&  \multicolumn{2}{c}{52.3}\\
Tukey's constant&  \multicolumn{2}{c|}{1.2}&  \multicolumn{2}{c|}{1.2}&  \multicolumn{2}{c|}{1.2}&  \multicolumn{2}{c}{1.2}
\\
\hline
\end{tabular}
\end{center}
\caption{Inside each class, the first two rows are the two pairs of gene expressions of the Khan dataset ordered by statistical depth with the least deep on the left and the second least deep on the right. The third row contains the distance between the two elements of each corresponding pair.  The fourth row is the value $G\cdot d(Q_3 ,Q_1)$ of Inequality \ref{mild3}, the benchmark to detect a pair of gene expressions as outliers. The fifth row is the value of $G$ from Table \ref{TK}. There are four  intra-pair distances and four corresponding gene expressions  in bold to represent i potential outliers with respect to their class.}\label{TKc}
\end{table}

\subsection{Tissue}
The Tissue data contains one outlier when applied to the whole set of data. This can be seen in Table \ref{TT} where we have  displayed the eight statistical least deep pairs of microarrays and the statistical least deep pair has a distance of $38.7$ between its two elements, that is larger than the outlier's benchmark of $38.3.$ In the graphical procedures, we can observe that microarray 34 is the one with highest OY-axis in the multidimensional scaling plot, see the bottom right plot of Figure \ref{A} for the multidimensional scaling procedure and the  bottom right plot of Figure 3 
of the Supplementary material for the spectral map analysis. However, it would not be clear whether to flag it as a potential outlier. It would be more clear, however, to point to microarray number 20 as a potential outlier in its class. However, although the analytical procedure contains it in the least deep microarray of its class, it does not detect it as potential outlier. In fact, the analytical procedure, detects no outlier when the interquartile distance is computed for each class separately, see Tables \ref{TTc}.

\begin{table}[!htb]
\begin{center}
\begin{tabular}{r|cccccccccc}
& \multicolumn{8}{c}{Tissue}
\\
\hline
&&&&&&&&
\\
pairs of gene &{\bf 34} &35& 41& 30& 32& 24&  5& 10 
\\
expressions &22 & 4&  8 & 9&  3&  2& 20& 17 
\\
&&&&&&&&
\\
distance intra-pair& {\bf38.7} &34.6 &32.3 &31.7 &30.5&30.4&29.6 &29.5
\\
&&&&&&&&\\
outlier's benchmark & \multicolumn{8}{c}{38.3}\\
Tukey's constant&  \multicolumn{8}{c}{1.3}\\
\hline
\end{tabular}
\end{center}
\caption{The first two rows are the eight statistical less deep pairs of gene expressions of the Tissie dataset, with the statistical least deep on the left. The third row contains the distance between the two elements of each corresponding pair.  The fourth row is the value $G\cdot d(Q_3 ,Q_1)$ of Inequality \ref{mild3}, the benchmark of the dataset to detect whether each pair of gene expressions are potential  outliers. The fifth row is the value of $G$ obtained for this dataset.  In bold it is the first intra-pair distance and the corresponding gene expression that is farther away in $L_2$ (euclidean) distance from the deepest gene expression.}\label{TT}
\end{table}

\begin{table}[!htb]
\begin{center}
\begin{tabular}{r|cc|cc|cc}
& \multicolumn{6}{c}{Tissue}
\\
& \multicolumn{2}{c}{class 1}& \multicolumn{2}{c}{class 2}& \multicolumn{2}{c}{class 3}
\\
\hline
&&&&&&
\\
pairs of gene & 8  &  1 &  20 &  22&  34&24
\\
expressions &  7 &  9& 13 & 19&  23&30
   \\
&&&&&&
\\
distance intra-pair& 26.5 &24.1&23.9 &23.4&36.2& 29.7
\\
&&&&&&\\
outlier's benchmark & \multicolumn{2}{c|}{30.4}&  \multicolumn{2}{c|}{40.9}&  \multicolumn{2}{c}{41.1}\\
Tukey's constant&  \multicolumn{2}{c|}{1.3}&  \multicolumn{2}{c|}{1.3}&  \multicolumn{2}{c}{1.3}\\
\hline
\end{tabular}
\end{center}
\caption{Inside each class, the first two rows are the two pairs of gene expressions of the Tissue dataset ordered by statistical depthwith the least deep on the left and the second least deep on the right. The third row contains the distance between the two elements of each corresponding pair.  The fourth row is the value $G\cdot d(Q_3 ,Q_1)$ of Inequality \ref{mild3}, the benchmark to detect a pair of gene expressions as outliers. The fifth row is the value of $G$ from Table \ref{TT}.}\label{TTc}
\end{table}

\section{A comparison with Robust Multi-array Average} 

Robust multi-array average, generally known as RMA, is a conventional pre-processing technique that makes use of quantile normalization. We develop here a corresponding technique that makes use of our statistical functional depth based normalization, and refer to it as \emph{Functional Depth Normalization (FDN)}. We compare both through simulations in what follows.

\subsection{Simulations}\label{simu}

In order to compare FDN with RMA, we performed a simulation involving 12 samples with 1000 genes and 11000 probes per sample. Each probe was generated from a t-distribution center at 3. We performed the procedure three times for the t-distribution with 10, 5 and 2 degrees of freedom. The few negative values were replaced by a small positive number. Then, the first 11000 probes of the first six samples were shifted by a positive $\delta$ taking values $\delta=0,0.25,0.5,1,2$  to create a difference in 100 genes. The other 900 genes should show no difference.
Finally the resulting probe values for each of the 12 samples were elevated to the power of $3+\epsilon$ where  the 12 epsilon values where sampled from a uniform(0,2) distribution. This last step applies a non-linear scale change to each sample separately, to create the need for normalization.

We generated 100 datasets following this method and we calculated gene expression sets using RMA including full median polish, not one-step median polish as was performed in the original RMA  \cite{Cabrera1, Irizarry}. The normalization step was performed using quantile normalization (RMA) or with statistical functional depth normalization (FDN). To study the performance of both methods we applied a conditional t-test \cite{AC2007} to compare the gene expressions between the two groups. Recall that the first 100 genes were differentially expressed by $\delta$ and the other 900 were not. The results are given in Table \ref{rma} columns 1 and 2 showing the number of true discoveries out the 100, which is equivalent to the power. It appears that for this example the FDN technique produces slightly better power with less false discoveries (see columns 4 and 5 of Table  \ref{rma}), but they are not very different. 
 In Table \ref{rma} column 3, and 6, we present the result of the simulation using FDN with Tukey's biweight M-estimator replacing median polish, for comparing FDN with Median Polish. Table \ref{rma} shows that for this simulation the power of the conditional t-test for FDN with the M-estimator is better than the power for FDN with median polish.
In conclusion this simulation shows situations where FDN together with Tukey's biweight M-estimator are good choices for implementing an alternative RMA algorithm.

\begin{table}[!htb]
\begin{center}
\begin{tabular}{cr|ccc|ccc}
&& \multicolumn{3}{c}{Power}&\multicolumn{3}{c}{False Discovery}
\\
&&&&&&&\\
degrees&&& Median& M-& &Median& M-\\
of&$\delta$&RMA&  Polish& Estimator& RMA&Polish& Estimator\\
freedom&& & FDN& FDN & & FDN& FDN 
\\
\hline
&0& {\color{blue}5.05}& 5.09&{\color{blue}{\bf 5.01}} &46.40&{\color{blue}44.86}& {\color{blue}{\bf44.63}} \\
&0.25& 15.25& {\color{blue}15.28}&{\color{blue}{\bf 19.41}} &{\color{blue}48.06}&48.30&{\color{blue}{\bf47.22 }} \\
10&0.5&  {\color{blue}45.01}&  44.99& {\color{blue}{\bf 56.33}}&52.25&{\color{blue}51.33}&{\color{blue}{\bf50.26}} \\
&1&   94.94& {\color{blue}95.03}&{\color{blue}{\bf 97.11}}&{\color{blue}{\bf69.51}} &{\color{blue}72.09}& 72.11\\
&2& {\color{blue}99.99} & {\color{blue}{\bf 100}}& {\color{blue}99.99} &147.59&{\color{blue}144.23}&{\color{blue}{\bf142.21}} \\
&&&&&&&\\
&0&5.31 &{\color{blue}5.15} & {\color{blue}{\bf 5.03 }}&{\color{blue}45.37}&45.46&{\color{blue}{\bf45.32}} \\
&0.25& 14.10& {\color{blue}14.59}& {\color{blue}{\bf 17.25}} &{\color{blue}45.92}&{\color{blue}{\bf44.42}} &46.38\\
5&0.5& 41.57& {\color{blue}41.86}& {\color{blue}{\bf 50.97}} &51.97&{\color{blue}{\bf50.12}} &{\color{blue}51.80}\\
&1& 93.11& {\color{blue}93.31}&{\color{blue} {\bf 97.22}} &68.00&{\color{blue}{\bf65.80}} &{\color{blue}66.53}\\
&2& {\color{blue}{\bf 100}}&{\color{blue}{\bf  100}}&  {\color{blue}99.98}&131.87&{\color{blue}{\bf127.31}} &{\color{blue}130.1}\\
&&&&&&&\\
&0&{\color{blue}{\bf  5.03}}& 5.12& {\color{blue}5.09}&{\color{blue}44.78}&{\color{blue}{\bf44.66}} &45.26\\
&0.25& 11.89& {\color{blue}12.38}&{\color{blue}{\bf  13.11}}&46.68&{\color{blue}{\bf45.49}} & {\color{blue}46.50}\\
2&0.5& {\color{blue}34.95}& 34.45&{\color{blue}{\bf   36.76}}&48.95&{\color{blue}{\bf46.04}} &{\color{blue}47.91}\\
&1& 85.26& {\color{blue}85.45}&{\color{blue}{\bf  88.66}} &59.20&{\color{blue}{\bf56.81}} &{\color{blue}57.16}\\
&2&{\color{blue}{\bf  99.98}}& {\color{blue}99.96}& {\color{blue} 99.96}&102.46&{\color{blue}{\bf99.10}} &{\color{blue}99.50}\\
\hline
\end{tabular}
\end{center}
\caption{Power results and false discoveries out of 900 of the conditional t-test for RMA (columns 1 and 4), FDN with median polish (column 2 and 5) and FDN with M-Estimator (column 3 and 6). The best result of the three in each case is in bold blue and the second best in blue.}\label{rma}
\end{table}

\section{Discussion and conclusion}

The propose methodology allows to perform a robust normalization procedure that satisfy the appropriate mathematical properties and an outlier detection methodology that is unambiguous, no depending on the expert pursuing it.

Our research has found the following pressure findings: 
\begin{itemize}
\item The Airway dataset contains one outlier and when restricting two each class separately, another outlier is found. The analytical procedures would have suggested, however, both RNA-seq's as outliers of the whole dataset.

\item The Sialin dataset after 6 hours of gestation detects the same outlier when studying the dataset as a whole and when studying it by classes. The graphical procedures suggest the same outlier and one additional outlier.

\item The Sialin dataset after 18 days contains one outlier when the data is study as a whole which is not spotted by the graphical procedures.
Like the Tissue data, below, it contains no outlier when  the data is studied by classes.

\item The Khan dataset has no outlier when studying the dataset as a whole. However, it contains outliers in three of its four clases when studying them separately. In fact, one of the clases contains two outliers.
\item In contraposition with the Khan dataset, the Tissue dataset contains an outlier  when studying the dataset as a whole but no outliers when studying the classes separately. The analytical procedure differs from the suggestion of the graphical procedures. For instance, the multidimensional scaling producere would suggest an outlier in the second class.
\end{itemize}
See Table \ref{out} for a summary.

\begin{table}[!htb]
\begin{center}
\begin{tabular}{r|ccccc}
& \multicolumn{5}{c}{Dataset} 
\\
&Airway&Sialin 6h.&Sialin 18 d.&Khan&Tissue\\
\hline 
Potential outliers id.& 5&3&3&-&34
\\
Potential outliers id. Class 1& -&-&-&16, 4&-
\\
Potential outliers id. Class 2& 6&3&-&-&-
\\
Potential outliers id. Class 3& na&na&na&32&-
\\
Potential outliers id. Class 4& na&na&na&51&na
\\
\hline
\end{tabular}
\end{center}
\caption{Summary of the potential outliers for the different datasets. The numbers in the table correspond to the RNA-seq or microarray identifier of Tables \ref{names1} and \ref{names2}. id. stands for identifier and NA stands for not aplicable.}\label{out}
\end{table}

\bibliographystyle{imsart-nameyear}	
\bibliography{Normalization}

\end{document}